\documentclass{article}
\usepackage{graphicx}
\usepackage{amsmath}

\newtheorem{theorem}{Theorem}

\newtheorem{definition}[theorem]{Definition}

\newtheorem{lemma}[theorem]{Lemma}

\newtheorem{proposition}[theorem]{Proposition}

\begin{document}

\title{Scattering Solutions in Networks of Thin Fibers: Small Diameter Asymptotics.}
\author{S. Molchanov, B. Vainberg \thanks{
The authors were supported partially by the NSF grant DMS-0405927.} \and %
Dept. of Mathematics, University of North Carolina at Charlotte, \and %
Charlotte, NC 28223, USA}
\date{}
\maketitle

\begin{abstract}
Small diameter asymptotics is obtained for scattering solutions in a network
of thin fibers. The asymptotics is expressed in terms of solutions of
related problems on the limiting quantum graph $\Gamma .$ We calculate the
Lagrangian gluing conditions at vertices $v\in \Gamma $ for the problems on
the limiting graph. If the frequency of the incident wave is above the
bottom of the absolutely continuous spectrum, the gluing conditions are
formulated in terms of the scattering data for each individual junction of
the network.
\end{abstract}


\bigskip

\section{Formulation of the problem and statement of the results}

The paper concerns the asymptotic analysis of wave propagation through a
system of wave guides when the thickness $\varepsilon $ of the wave guides
is very small and the wave length is comparable to $\varepsilon $. The
problem is described by the stationary wave (Helmholtz) equation
\begin{equation}
-\varepsilon ^{2}\Delta u=\lambda u,\text{ \ \ \ }x\in \Omega _{\varepsilon
},  \label{h0}
\end{equation}
in a domain $\Omega _{\varepsilon }\subset R^{d},$ $d\geq 2,$ with
infinitely smooth boundary (for simplicity) which has the following
structure: $\Omega _{\varepsilon }$ is a union of a finite number of
cylinders $C_{j,\varepsilon }$ (which we shall call channels)$,$ $1\leq
j\leq N,$ of lengths $l_{j}$ with the diameters of cross-sections of order $%
O\left( \varepsilon \right) $ and domains $J_{1,\varepsilon },\cdots
,J_{M,\varepsilon }$ (which we shall call junctions) connecting the channels
into a network. It is assumed that the junctions have diameters of the same
order $O(\varepsilon )$. Let $m$ channels have infinite length. We start the
numeration of $C_{j,\varepsilon }$ with the infinite channels. So, $%
l_{j}=\infty $ for $1\leq j\leq m.$ The axes of the channels form edges $%
\Gamma _{j}$ of the limiting $\left( \varepsilon \rightarrow 0\right) $
metric graph $\Gamma $. The vertices $v_j \in V$ of the graph $\Gamma $
correspond to the junctions $J_{j,\varepsilon }$.

\begin{figure}[htbp]
\begin{center}
\includegraphics[width=0.8\columnwidth]{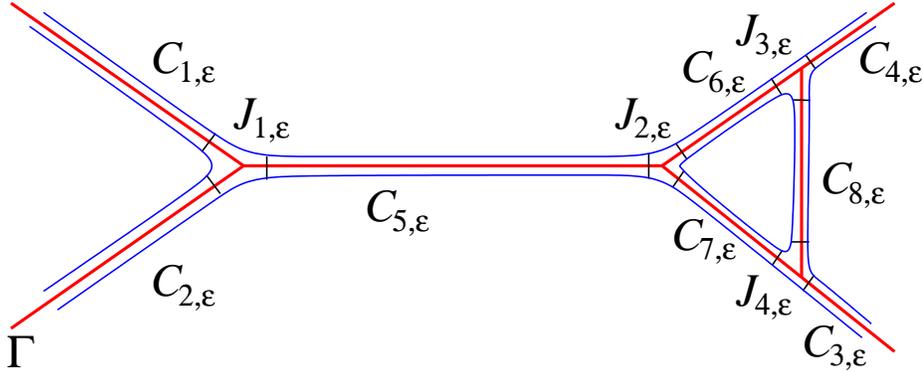}
\end{center}
\caption{An example of a domain $\Omega _{\protect\varepsilon }$ with four
junctions, four unbounded channels and four bounded channels.}
\label{fig-1}
\end{figure}

The Helmholtz equation in $\Omega _{\varepsilon }$ must be complemented by
the boundary conditions (BC) on $\partial \Omega _{\varepsilon }$. In some
cases (for instance, when studying heat transport in $\Omega _{\varepsilon }$%
) the Neumann BC is natural. In fact, the Neumann BC presents the simplest
case due to the existence of a simple ground state (a constant) of the
problem in $\Omega _{\varepsilon }.$ \ However, in many applications, the
Dirichlet, Robin or impedance BC are more important. We shall consider
(apart from a general discussion) only the Dirichlet BC, but all the
arguments and results can be modified to be applied to the problem with
other BC.

An important class of domains $\Omega _{\varepsilon }$ are self-similar
domains with only one junction and all the channels being infinite. We will
call them \textit{spider domains}. Thus, if $\Omega _{\varepsilon }$ is a
spider domain, then there exist a point $\widehat{x}=x(\varepsilon )$ and an
$\varepsilon $-independent domain $\Omega $ such that

\begin{equation}
\Omega _{\varepsilon }=\{(\widehat{x}+\varepsilon x):x\in \Omega \}.
\label{hom}
\end{equation}

Thus, $\Omega _{\varepsilon }$ is $\varepsilon $-contraction of $\Omega
=\Omega _{1}.$

For any $\Omega _{\varepsilon },$ let $J_{j(v),\varepsilon }$ be the
junction which corresponds to a vertex $v\in V$ of the limiting graph $%
\Gamma .$ Consider a junction $J_{j(v),\varepsilon }$ and all adjacent to $%
J_{j(v),\varepsilon }$ channels. If some of these channels have a finite
length, we extend them to infinity. We assume that, for each $v\in V,$ the
resulting domain $\Omega _{v,\varepsilon }$ which consists of junction $%
J_{j(v),\varepsilon }$ and emanating from it semi-infinite channels is a
spider domain (i.e., $\Omega _{v,\varepsilon }$ is self-similar). This
assumption can be weakened. For example, one can consider some type of
''curved'' channels, and the final results (with some changes) will remain
valid. Simple equations on the limiting graph in this case will be replaced
by more complicated equations with variable coefficients. However, even
small deviation from the assumption on the self-similarity of $\Omega
_{v,\varepsilon }$ would make the statement of the results and the proofs
much more technical. So, we consider only domains $\Omega _{\varepsilon }$
for which $\Omega _{v,\varepsilon },$ $v\in V,$ are self-similar.

Hence, the cross sections $\omega _{j,\varepsilon }\ $of channels $%
C_{j,\varepsilon }$ are $\varepsilon -$homothety of bounded domains $\omega
_{j}\in R^{d-1}$. \ Let $\lambda _{j,0}<\lambda _{j,1}\leq \lambda _{j,2}...$
be eigenvalues of the negative Laplacian $-\Delta _{d-1}$ in $\omega _{j}$
with the Dirichlet boundary condition on $\partial \omega _{j},$ and let $%
\{\varphi _{j,n}\}$ be the set of corresponding orthonormal eigenfunctions.\
The eigenvalues $\lambda _{j,n}$ coincide with the eigenvalues of $%
-\varepsilon ^{2}\Delta _{d-1}$ in $\omega _{j,\varepsilon }.$ $\ $In the
presence of infinite channels, the spectrum of the operator $-\varepsilon
^{2}\Delta $ in $\Omega _{\varepsilon }$ with the Dirichlet boundary
condition on $\partial \Omega _{\varepsilon }$ has an absolutely continuous
component which coincides with the semi-bounded interval $[\lambda
_{0},\infty ),$ where
\begin{equation}
\lambda _{0}=\min_{1\leq j\leq m}\lambda _{j,0}  \label{b1}
\end{equation}
The equation (\ref{h0}) is considered under the assumption that $\lambda
\geq \lambda _{0},$ when propagation of waves is possible. There are two
very different cases: $\lambda \rightarrow \lambda _{0}\ $as $\varepsilon
\rightarrow 0,$ i.e. the frequency is at the edge (or bottom) of the
absolutely continuous spectrum, or $\lambda \rightarrow \widehat{\lambda }%
>\lambda _{0},$ i.e. the frequency is above the bottom of the absolutely
continuous spectrum. There are many results about the first case, the
references will be given later. This paper concerns the asymptotic analysis
of the scattering solutions for the Dirichlet problem in $\Omega
_{\varepsilon }$ when $\lambda $ is close to $\widehat{\lambda }>\lambda
_{0}.$

If $\varepsilon \rightarrow 0$, one can expect that the solution $%
u_{\varepsilon }$ of (\ref{h0}) in $\Omega _{\varepsilon }$ can be described
in terms of the solution $\varsigma =\varsigma _{\varepsilon }(t)$ of a much
simpler problem on the graph $\Gamma $. For example, if $\lambda
_{j,0}<\lambda <\lambda _{j,1}$ for all $j$, then $\varsigma $ satisfies the
following equation on each edge of the graph
\begin{equation}
-\dfrac{\varepsilon ^{2}d^{2}\varsigma \left( t\right) }{dt^{2}}=(\lambda
-\lambda _{j,0})\varsigma \left( t\right) ,  \label{lsg}
\end{equation}
where $t$ is the length parameter on the edges. One has to add appropriate
gluing conditions (GC) at the vertices $v$ of $\Gamma $. These gluing
conditions give basic information on the propagation of waves through the
junctions. They define the solution $\varsigma $ of the problem (\ref{lsg})
on the limiting graph. The ordinary differential equation (\ref{lsg}), the
GC, and the solution $\varsigma $ depend on $\varepsilon .$ However, we
shall often call the corresponding problem on the graph the limiting
problem, since it enables one to find the main term of the asymptotics as $%
\varepsilon \rightarrow 0$ for the solution $u=u_{\varepsilon }$ of the
problem (\ref{h0}) in $\Omega _{\varepsilon }.$

One of the main difficulties in the problem under investigation was to find
the GC, in particular, since the GC differ dramatically from those which
were known in the case of $\lambda $ close to the bottom of the spectrum.

Let us define the scattering solutions for the Dirichlet problem in $\Omega
_{\varepsilon }.$ We introduce local coordinates $(t,y)$ in each channel $%
C_{j,\varepsilon }$\ with $t$ axis parallel to the cylinder $%
C_{j,\varepsilon },$ $0<t<l_{j},$\ and $y\in R^{n-1}$ being Euclidean
coordinates in the plane perpendicular to the $t$ axis. The coordinate $y$
is chosen in such a way that $\omega _{j,\varepsilon }=\{(\varepsilon
y):y\in \omega _{j}\in R^{n-1}\}.$ For each $j,$ the set $\{\varepsilon ^{%
\frac{1-d}{2}}\varphi _{j,n}(\frac{y}{\varepsilon })\}$ is the orthonormal
basis in $L^{2}(\omega _{j,\varepsilon })$ consisting of eigenfunctions of
the operator $-\varepsilon ^{2}\Delta _{d-1}.$

Let $l$ be a bounded closed interval of the real axis which does not contain
the points $\lambda _{j,n},$ $j\leq N.$ Thus, there exist $m_{j}\geq 1$ such
that $\lambda _{j,m_{j}}<\lambda <\lambda _{j,m_{j}+1}$ for all $\lambda \in
l.$ As will be seen from the definitions below, $m_{j}+1$ is the number of
waves which may propagate in each direction in the channel $C_{j,\varepsilon
}$ without loss of energy and with frequencies less than $\sqrt{\lambda },$ $%
\lambda \in l$ . We put $m_{j}=-1$, thus $\{\lambda _{j,n},$ $0\leq n\leq
m_{j}\}$ is the empty set if $\lambda _{j,0}>\lambda $ for $\lambda \in l.$

Consider the non-homogeneous Dirichlet problem
\begin{equation}
(-\varepsilon ^{2}\Delta -\lambda )u=f,\text{ \ }x\in \Omega _{\varepsilon };%
\text{ \ \ \ }u=0\text{ \ on }\partial \Omega _{\varepsilon }.  \label{a1}
\end{equation}

\begin{definition}
\label{d1}Let $f\in L_{com}^{2}(\Omega _{\varepsilon })$ have a compact
support, and $\lambda \in l$.\ A solution $u$ of (\ref{a1}) is called an
outgoing solution if it has the following asymptotic behavior at infinity in
each infinite channel $C_{j,\varepsilon },$ \ \ $1\leq j\leq m$:
\begin{equation}
u=\sum_{n=0}^{m_{j}}a_{j,n}e^{i\frac{\sqrt{\lambda -\lambda _{j,n}}}{%
\varepsilon }t}\varphi _{j,n}(y/\varepsilon )+O(e^{-\gamma t}),\text{ \ \ }%
\gamma =\gamma (\varepsilon )>0,  \label{a2}
\end{equation}

\begin{definition}
\label{d2}A function $\Psi =\Psi _{s,k}^{(\varepsilon )},$ $1\leq s\leq m,$ $%
0\leq k\leq m_{j},$ is called a solution of the scattering problem in $%
\Omega _{\varepsilon }$ if
\begin{equation}
(-\varepsilon ^{2}\Delta -\lambda )\Psi =0,\text{ \ }x\in \Omega
_{\varepsilon };\text{ \ \ \ }\Psi =0\text{ \ on }\partial \Omega
_{\varepsilon },  \label{b9}
\end{equation}
and $\Psi $ has the following asymptotic behavior at infinity in each
infinite channel $C_{j,\varepsilon },$ \ \ $1\leq j\leq m$:
\begin{equation}
\Psi _{s,k}^{(\varepsilon )}=\delta _{s,j}e^{-i\frac{\sqrt{\lambda -\lambda
_{s,k}}}{\varepsilon }t}\varphi _{s,k}(y/\varepsilon
)+\sum_{n=0}^{m_{j}}t_{j,n}e^{i\frac{\sqrt{\lambda -\lambda _{j,n}}}{%
\varepsilon }t}\varphi _{j,n}(y/\varepsilon )+O(e^{-\gamma t}),  \label{b10}
\end{equation}
where $\gamma =\gamma (\varepsilon )>0,$ and $\delta _{s,j}$ is the
Kronecker symbol, i.e. $\delta _{s,j}=1$ if $s=j,$ $\delta _{s,j}=0$ if $s$ $%
\neq j.$
\end{definition}
\end{definition}

The first term in (\ref{b10}) corresponds to the incident wave, and all
other terms describe the transmitted waves. The incident wave depends on $s$
and $k,$ where $s$ determines the channel, and $s$ and $k$ together
determine the frequency of the incident wave. The transmission coefficients $%
t_{j,n}$ also depend on $s$ and $k$ (i.e. on the choice of the incident wave)%
$,$ so sometimes we will denote them by $t_{j,n}^{s,k}.$

We introduce an order in the set of incident waves and corresponding
scattering solutions and the same order in the set of transmitted waves.
Namely, we number the incident waves in the channel $C_{1,\varepsilon }$
taking them in the order of increase of absolute values of their
frequencies, then we number all the solutions in the channel $%
C_{2,\varepsilon },$ and so on. With this order taken into account, the
transmission coefficients for a particular scattering solution form a column
vector with
\begin{equation}
M=\sum_{j=1}^{m}(m_{j}+1)  \label{em}
\end{equation}
entries$.$ Together, they form an $M\times M$ \textit{scattering matrix}
\begin{equation}
T=\{t_{j,n}^{s,k}\},  \label{scm}
\end{equation}
where $s,k$ define the column of $T$ and $j,n$ define the row. We denote by $%
D$ the diagonal $M\times M$ matrix with elements $\sqrt{\lambda -\lambda
_{j,n}}$ on the diagonal taken in the same order as above. The following
statement can be useful in some applications, and will be proved in the next
section (although it will not be used in this paper).

\begin{theorem}
\label{t3}The matrix $D^{1/2}TD^{-1/2}$ is unitary and symmetric.
\end{theorem}

The operator $H=-\varepsilon ^{2}\Delta $ with the Dirichlet boundary
conditions on $\partial \Omega _{\varepsilon }$ is non-negative, and
therefore the resolvent
\begin{equation}
R_{\lambda }=(-\varepsilon ^{2}\Delta -\lambda )^{-1}:L^{2}(\Omega
_{\varepsilon })\rightarrow L^{2}(\Omega _{\varepsilon })  \label{res}
\end{equation}
is analytic in the complex $\lambda $ plane outside the positive semi-axis $%
\lambda \geq 0.$ Hence, the operator $R_{k^{2}}$ is analytic in $k$ in the
half plane Im$k>0.$ We are going to consider an analytic extension of the
operator $R_{k^{2}}$ onto the real axis and in the lower half plane. Such an
extension does not exist if $R_{k^{2}}$ is considered as an operator in $%
L^{2}(\Omega _{\varepsilon })$ since $R_{k^{2}}$ is an unbounded operator
when $\lambda =k^{2}$ belongs to the spectrum of the operator $R_{\lambda }.$
However, one can extend $R_{k^{2}}$ analytically if it is considered as an
operator in the following spaces (with a smaller domain and a larger range):
\begin{equation}
R_{k^{2}}:L_{com}^{2}(\Omega _{\varepsilon })\rightarrow L_{loc}^{2}(\Omega
_{\varepsilon }).  \label{b2}
\end{equation}

\begin{theorem}
\label{t1}(1) The spectrum of the operator $H=-\varepsilon ^{2}\Delta $ in $%
\Omega _{\varepsilon }$ with the Dirichlet boundary conditions on $\partial
\Omega _{\varepsilon }$ consists of the absolutely continuous component $%
[\lambda _{0},\infty )$ where $\lambda _{0}>0$ is given by (\ref{b1}) and,
possibly, a discrete set of positive eigenvalues $\{\lambda ^{j,\varepsilon
}\}$ with the only possible limiting point at infinity.\ The multiplicity of
the a.c. spectrum changes at points $\lambda =\lambda _{j,n},$ and at any
point $\lambda ,$ it is equal to the number of points $\lambda _{j,n},$ $%
1\leq j\leq m,$\ located below $\lambda .$ The eigenvalues $\lambda
^{j,\varepsilon }=\lambda ^{j}$ for spider domains $\Omega _{\varepsilon }$
do not depend on $\varepsilon .$

(2) The operator (\ref{b2}) admits a meromorphic extension from the upper
half plane Im$k>0$ into lower half plane Im$k<0$ with the branch points at $%
k=\pm \sqrt{\lambda _{j,n}}$ of the second order and the real poles at $%
k=\pm \sqrt{\lambda ^{s,\varepsilon }}$ and, perhaps, at some of the branch
points. The resolvent (\ref{b2}) has a pole at $k=\pm \sqrt{\lambda _{j,n}}$
if and only if the homogeneous problem (\ref{a1}) with $\lambda =\lambda
_{j,n}$ has a nontrivial solution $u$ such that
\begin{equation}
u=\sum_{j,n:\lambda _{j,n}=\lambda }a_{j,n}\varphi _{j,n}(y/\varepsilon
)+(e^{-\gamma t}),\text{ \ \ \ }x\in C_{j,\varepsilon },\text{ \ \ }%
t\rightarrow \infty ,\text{ \ \ }1\leq j\leq m.  \label{inf}
\end{equation}

(3) If $f\in L_{com}^{2}(\Omega _{\varepsilon }),$ and $k=\sqrt{\lambda }$
is real and is not a pole or a branch point of \ the operator (\ref{b2}),
and $\lambda >\lambda _{0},$ then the problem (\ref{a1}), (\ref{a2}) is
uniquely solvable and the outgoing solution $u$ can be found as the $%
L_{loc}^{2}(\Omega _{\varepsilon })$ limit
\begin{equation}
u=R_{\lambda +i0}f.  \label{b3}
\end{equation}

(4) There exist exactly $M\ $(see (\ref{em})) different scattering solutions
for values of $\lambda >\lambda _{0}$ such that $k=\sqrt{\lambda }$ is not a
pole or a branch point of \ the operator (\ref{b2}), and the scattering
solution is defined uniquely after the incident wave is chosen.
\end{theorem}

\textbf{Remarks. }1. Operator $H=-\varepsilon ^{2}\Delta $ and its domain
depend on $\varepsilon .$ One could use the term ''family of operators''
when referring to $H$. We prefer to drop the word ''family'', but one must
always keep in mind that $H$ depends on $\varepsilon .$

2.\textbf{\ }Existence of a pole\ of the operator (\ref{b2}) at a branch
point means that $R_{k^{2}}$ has a pole at $z=0$ if this operator function
is considered as a function of $z=\sqrt{k^{2}-\lambda _{j,n}}.$

3. One can not identify poles of the resolvent and eigenvalues of the
operator based only on general theorems of functional analysis since we deal
with the poles of the modified resolvent (\ref{b2}) which belong to the
absolutely continuous spectrum of the operator.

4. The eigenvalues $\lambda ^{j,\varepsilon }$ of the operator $H$ can be
embedded into the absolutely continuous spectrum, and can be located below
the absolutely continuous spectrum. In particular, from the minimax
principle it follows that $H$ necessarily has a non-empty discrete spectrum
below $\lambda _{0}$ if at least one of the junctions is wide enough. For
example, non-empty discrete spectrum below $\lambda _{0}$ exists if a
junction contains a ball $B_{\rho }$ of the radius $\rho =r\varepsilon $
such that the negative Dirichlet Laplacian in the ball $B_{r}$ has an
eigenvalue below $\lambda _{0}.$

Let us describe the asymptotic behavior of scattering solutions $\Psi =\Psi
_{s,k}^{(\varepsilon )}$ as $\varepsilon \rightarrow 0,$ $\lambda \in l.$
Note that an arbitrary solution $u$ of the equation (\ref{h0}) in a channel $%
C_{j,\varepsilon }$ can be represented as a series with respect to the
orthogonal basis $\{\varphi _{j,n}(y/\varepsilon )\}$ of the eigenfunctions
of the Laplacian in the cross-section of $C_{j,\varepsilon }.$ Thus it can
be represented as a linear combination of the travelling waves
\begin{equation*}
e^{\pm i\frac{\sqrt{\lambda -\lambda _{j.n}}}{\varepsilon }t}\varphi
_{j,n}(y/\varepsilon ),\text{ \ \ }1\leq n\leq m_{j},
\end{equation*}
and functions which grow or decay exponentially along the axis of $%
C_{j,\varepsilon }.$ The main term of small $\varepsilon $ asymptotics of
scattering solutions contains only travelling waves, i.e. on each channel $%
C_{j,\varepsilon },$ any function $\Psi $ has the form
\begin{equation}
\Psi =\Psi _{s,k}^{(\varepsilon )}=\sum_{n=0}^{m_{j}}(\alpha _{j,n}e^{i\frac{%
\sqrt{\lambda -\lambda _{j.n}}}{\varepsilon }t}+\beta _{j,n}e^{-i\frac{\sqrt{%
\lambda -\lambda _{j.n}}}{\varepsilon }t})\varphi _{j,n}(y/\varepsilon
)+r_{s,k}^{\varepsilon },  \label{psias}
\end{equation}
where
\begin{equation*}
|r_{s,k}^{\varepsilon }|\leq Ce^{-\frac{\gamma d(t)}{\varepsilon }},\text{ \
\ }\gamma >0,\text{ \ \ and \ }d(t)=\min (t,l_{j}-t).
\end{equation*}
The constants $\alpha _{j,n}$ and $\beta _{j,n}$ depend also on $s,k$ and $%
\varepsilon .$ The formula (\ref{psias}) can be written in a shorter form as
follows
\begin{equation*}
\Psi =\Psi _{s,k}^{(\varepsilon )}=\sum_{n=0}^{m_{j}}\varsigma _{j}\cdot
\varphi _{j}+r_{s,k}^{\varepsilon },\text{ \ \ \ }|r_{s,k}^{\varepsilon
}|\leq Ce^{-\frac{\gamma d(t)}{\varepsilon }},
\end{equation*}
where $\varphi _{j}=\varphi _{j}(y/\varepsilon )$ is the vector with
components $\varphi _{j,n}(y/\varepsilon ),$ $0\leq n\leq m_{j},$ and $%
\varsigma _{j}=\varsigma _{j}(t)$ is a $(m_{j}+1)$-vector whose components $%
\varsigma _{j,n}$ are linear combinations of the corresponding oscillating
exponents in $t$, i.e. $\varsigma _{j}$ satisfies the following equation:
\begin{equation}
(\varepsilon ^{2}\frac{d^{2}}{dt^{2}}+D_{j}^{2})\varsigma _{j}=0,\text{ \ \ }%
0<t<l_{j},  \label{greq}
\end{equation}
where $D_{j}$ is the diagonal matrix with elements $\sqrt{\lambda -\lambda
_{j.n}},$ $0\leq n\leq m_{j},$ on the diagonal.

In order to complete the description of the main term of the asymptotic
expansion (\ref{psias}), we need to provide the choice of constants in the
representation of $\varsigma _{j,n}$ as linear combinations of the
exponents. Thus, $2(m_{j}+1)$ constants must be chosen for each channel $%
C_{j,\varepsilon }.$ We consider the limiting graph $\Gamma ,$ whose edges $%
\Gamma _{j}$ are the axes of the channels $C_{j,\varepsilon }.$ Let $%
\varsigma $ be the vector valued function on $\Gamma $ which is equal to $%
\varsigma _{j}$ on $\Gamma _{j}.$ The vector $\varsigma $ has a different
number of coordinates on different edges $\Gamma _{j}$ of the graph $\Gamma
. $ We specify $\varsigma $ by imposing conditions at infinity and gluing
conditions (GC) at each vertex $v$ of the graph $\Gamma .$ Let $V=\{v\}$ be
the set of vertices $v$ of the limiting graph $\Gamma .$ These vertices
correspond to the junctions in $\Omega _{\varepsilon }.$

The conditions at infinity concern only the infinite channel $%
C_{j,\varepsilon },$ $j\leq m.$ They depend on the choice of the incident
wave and have the form:
\begin{equation}
\beta _{j,n}=\left\{
\begin{array}{c}
1\text{ if \ }(j,n)=(s,k) \\
0\text{ if \ }(j,n)\neq (s,k)
\end{array}
\right. ,\text{ \ }1\leq j\leq m.  \label{uv}
\end{equation}

The GC at vertices $v$ of the graph $\Gamma $ are universal for all incident
waves and depend on $\lambda $. In order to state the GC at a vertex $v$ ,
we choose the parametrization on $\Gamma $ in such a way that $t=0$ at $v$
for all edges adjacent to this particular vertex$.$ The origin ($t=0$) on
all other edges can be chosen at any of the end points of the edge. Consider
auxiliary scattering problems for the spider type domain $\Omega
_{v,\varepsilon }$ formed by the individual junction, which corresponds to
the vertex $v,$ and all channels with an end at this junction, where the
channels are extended to infinity if they have a finite length. We denote by
$\Gamma _{v}$ the limiting graph which is defined by $\Omega _{v,\varepsilon
}.$ Definitions \ref{d1}, \ref{d2} and Theorem \ref{t1} remain valid for the
domain $\Omega _{v,\varepsilon }.$ In particular, one can define the
scattering matrix $T=T_{v}$ for the problem (\ref{h0}) in the domain $\Omega
_{v,\varepsilon }.$ Let $v_{1},$ $v_{2},...v_{l},$ $l=l(v),$\ be indices of
channels in $\Omega _{\varepsilon }$ which correspond to channels in $\Omega
_{v,\varepsilon }.$ Les us form a vector $\varsigma ^{(v)}$ by writing the
coordinates of all vectors $\varsigma _{v_{s}}$ in one column, starting with
coordinates of $\varsigma _{v_{1}},$ then coordinates of $\varsigma _{v_{2}},
$ and so on. Let us denote by $D_{v}(\lambda )$ the diagonal matrix with the
diagonal elements $\sqrt{\lambda -\lambda _{v_{s,k}}}$ written in the same
order as the coordinates of the vector $\varsigma ^{(v)}.$ Let $I_{v}$ be
the unit matrix of the same size as the size of the matrix $D_{v}(\lambda ).$
The GC at the vertex $v$ has the form
\begin{equation}
\varepsilon \lbrack I_{v}+T_{v}]D_{v}^{-1}(\lambda )\frac{d}{dt}\varsigma
^{(v)}(t)+i[I_{v}-T_{v}]\varsigma ^{(v)}(t)=0,\text{ \ \ \ }t=0.  \label{gc}
\end{equation}
The GC (\ref{gc}) has the following form in the coordinate representation.
Let $Z=Z(v)$ be the set of indices $(j,n),$ where $j$ are the indices of the
edges of $\Gamma $ ending at $v$ and $0\leq n\leq m_{j}.$ Then
\begin{eqnarray*}
\sum_{(j,n)\in Z}\left\{ \varepsilon \left[ \delta
_{j,n}^{s,k}+t_{j,n}^{s,k}(v)\right] (\lambda -\lambda _{j,n})^{-1/2}\frac{d%
}{dt}\varsigma _{j,n}+i\left[ \delta _{j,n}^{s,k}-t_{j,n}^{s,k}(v)\right]
\varsigma _{j,n}\right\}  &=&0\text{ \ at \ }v, \\
\text{ \ \ }(s,k) &\in &Z,\text{ }
\end{eqnarray*}
\ where $t_{j,n}^{s,k}(v)$ are the transmission coefficients of the
auxiliary problem in the spider domain $\Omega _{v,\varepsilon }$ (i.e. $%
t_{j,n}^{s,k}(v)$ are the elements of $T_{v}$),\ and $\delta _{j,n}^{s,k}=1$
if$\ (s,k)=(j,n),$ $\delta _{j,n}^{s,k}=0$ if$\ (s,k)\neq (j,n).$

\begin{definition}
A family of subsets $l(\varepsilon )$ of a bounded closed interval $l\subset
R^{1}$ will be called thin if, for any $\delta >0,$ there exist\ constants $%
\beta >0$ and $c_{1},$ independent of $\delta $ and $\varepsilon ,$ and $%
c_{2}=$ $c_{2}(\delta ),$ such that $l(\varepsilon )$ can be covered by $%
c_{1}$ intervals of length $\delta $ together with $c_{2}\varepsilon ^{-1}$
intervals of length $c_{2}e^{-\beta /\varepsilon }.$ Note that $%
|l(\varepsilon )|\rightarrow 0$ as $\varepsilon \rightarrow 0.$
\end{definition}

\begin{theorem}
\label{t2}Let $l$ be a bounded closed interval of the $\lambda $-axis which
does not contain points $\lambda _{j,n}.$ Then there exists $\gamma =\gamma
(\omega _{j},l)>0$ and a thin family of sets $l(\varepsilon )$ such that the
asymptotic expansion (\ref{psias}) holds on all (finite and infinite)
channels $C_{j,\varepsilon }$ uniformly in $\lambda \in l$ $\backslash $ $%
l(\varepsilon )$ and $x$ in any bounded region of $R^{d}.$ The function $%
\varsigma $ in (\ref{psias}) is a vector function on the limiting graph
which satisfies equation (\ref{greq}), conditions (\ref{uv}) at infinity,
and the GC (\ref{gc}).
\end{theorem}

\textbf{Remarks}. 1) It will be shown in the proof of Lemma \ref{l21} that
for spider domains the estimate of the remainder is uniform for all $x\in
R^{d}$. For general domains, we provide the estimate of the remainder only
in bounded regions of $R^{d}$ in order not to complicate the exposition$.$

2) The arguments, used to justify the asymptotic behavior of the scattering
solutions and prove Theorem \ref{t2}, can be applied to the study the
asymptotic behavior of the outgoing solutions of the non-homogeneous problem
(\ref{a1}) as $\varepsilon \rightarrow 0,$ $\lambda >\lambda _{0}.$ The
asymptotics will be expressed in terms of solutions of the corresponding
non-homogeneous equation on the limiting graph. One can easily show that the
GC can not be chosen independently of $f$ even if we consider only functions
$f$ with compact support. However, if the support of $f$ is separated from
the junctions then the solution of the non-homogeneous equation on the
limiting graph satisfies the same universal GC (\ref{gc})\ that appear when
scattering solutions are studied. The latter is related to the following
fact: the outgoing solution in a narrow channel behaves as a combination of
plane waves plus a term which decays exponentially outside of the support of
$f$ when $\varepsilon \rightarrow 0.$

Note that the GC for the function $\varsigma $ on the limiting graph depend
on $\lambda .$ In fact, there exists an effective matrix potential on $%
\Gamma $ which is independent of $\lambda ,$ and allows one to single out
the scattering solutions $\varsigma $ on $\Gamma $ with the same scattering
data as for the original problem in $\Omega _{\varepsilon }.$ These results
will be published elsewhere.

The convergence of the spectrum of the problem in $\Omega _{\varepsilon }$
to the spectrum of a problem on the limiting graph has been extensively
discussed in physical and mathematical literature (e.g., \cite{DE}-\cite{EP}%
, \cite{K,KZ1,KZ2,P,RS} and references therein). What makes our paper
different is the following: all the publications that we are aware of, are
devoted to the convergence of the spectra (or resolvents) only in a small
(in fact, shrinking with $\varepsilon \rightarrow 0)$ neighborhood of $%
\lambda _{0}$ (bottom of the absolutely continuous spectrum), or below $%
\lambda _{0}$. Usually, the Neumann BC on $\partial \Omega _{\varepsilon }$
is assumed. We deal with asymptotic behavior of solutions of the scattering
problem in $\Omega _{\varepsilon }$ when $\lambda $ is close to $\widehat{%
\lambda }>\lambda _{0},$ and the BC on $\partial \Omega _{\varepsilon }$ can
be arbitrary.

In particular, papers \cite{FW}, \cite{KZ1}, \cite{KZ2}, \cite{RS} contain
the gluing conditions and the justification of the limiting procedure $%
\varepsilon \rightarrow 0$ near the bottom of the spectrum $\lambda _{0}$
under assumption that the Neumann BC is imposed at the boundary of $\Omega
_{\varepsilon }.$ Note that $\lambda _{0}=0$ for the Neumann BC. Typically,
the GC in this case are:$\;$the continuity of $\varsigma \left( s\right) $
at each vertex $v$ and $\sum_{j=1}^{d}\varsigma _{j}^{\prime }\left(
v\right) =0$, i.e. the continuity of both the field and the flow. These GC
are called Kirchhoff's GC. In the case when the shrinkage rate of the volume
of the junction neighborhoods is lower than the one of the area of the
cross-sections of the guides, more complex energy dependent or decoupling
condition can arise (see \cite{K}, \cite{KZ2}, \cite{EP} for details). Let
us stress again that this is the situation near the bottom $\lambda _{0}=0$
of the absolutely continuous spectrum. As follows from Theorem \ref{t3}, the
GC and the small $\varepsilon $ asymptotics are different when $\widehat{%
\lambda }>\lambda _{0}.$

Both assumptions ($\lambda \rightarrow \lambda _{0},$ and the fact that the
BC is the Neumann condition) in the papers above are very essential. The
Dirichlet Laplacian near the bottom of the absolutely continuous spectrum $%
\lambda _{0}>0$ was studied in a recent paper \cite{P} under the condition
that the junctions are more narrow than the tubes. It is assumed there that
the domain $\Omega _{\varepsilon }$ is bounded. Therefore, the spectrum of
the operator (\ref{h0}) is discrete. It is proved that the eigenvalues of
the operator (\ref{h0}) in $O(\varepsilon ^{2})$-neighborhood of $\lambda
_{0}$ behave asymptotically, when $\varepsilon \rightarrow 0,$ as
eigenvalues of the problem in the disconnected domain that one gets by
omitting the junctions, separating the channels in $\Omega _{\varepsilon },$
and adding the Dirichlet conditions on the bottoms of the channels. This
result indicates that the waves do not propagate through the narrow
junctions when $\lambda $ is close to the bottom of the absolutely
continuous spectrum. A similar result was obtained in \cite{IT} for the Schr%
\"{o}dinger operator with a potential having a deep strict minimum on the
graph, when the width of the walls shrinks to zero.

We also studied the Dirichlet problem for general domains $\Omega
_{\varepsilon }$ without special assumptions on the geometry of the
junctions when, simultaneously, $\varepsilon \rightarrow 0,$ $\lambda
\rightarrow \lambda _{0},$ and the diameters of the guides and junctions
have the same order $O(\varepsilon ).$ Our conclusion is that, generically,
waves do not propagate through the junctions when the frequency is close to
the bottom of the absolutely continuous spectrum. Let us stress that this is
true both in the case when the diameters of the junctions are smaller than
the diameters of the guides, and in the case when they are larger. Some
special conditions must be satisfied for waves to propagate if $\lambda
\rightarrow \lambda _{0}$. An infinite cylinder, which can be considered as
two half-infinite tubes with the junction of the same shape, can be
considered as an example of a domain where the propagation of waves at $%
\lambda =\lambda _{0}$ is not suppressed. Less trivial examples will be
given in our next paper. We do not deal with the problem near the bottom of
the absolutely continuous spectrum in this publication. A detailed analysis
of this problem will be published elsewhere. However, we show here that the
GC on the limiting graph with $\lambda >\lambda _{0}$, generically, have a
limit as $\varepsilon \rightarrow 0,$ $\lambda \rightarrow \lambda _{0},$
and the limiting conditions are the Dirichlet conditions. To be more exact,
the following statement will be proved.

\begin{theorem}
\label{tlas}1) Assume that the resolvent (\ref{b2}) does not have a pole at $%
k=\sqrt{\lambda _{0}}.$ Then the scattering matrix (\ref{scm}), defined for $%
\lambda >\lambda _{0},$ admits an analytic in $z=\sqrt{\lambda -\lambda _{0}}
$ extension to a neighborhood of the point $z=0$ and is equal to $-I$ at $%
z=0,$ where $I$ is the $(m_{0}\times m_{0})$-identity matrix and $m_{0}$ is
the number of infinite channels $C_{j,\varepsilon }$ with $\lambda
_{j,0}=\lambda _{0}.$

2) Assume that the resolvent of the auxiliary problem in the spider type
domain $\Omega _{\nu ,\varepsilon }$ does not have a pole at $k=\sqrt{%
\lambda _{0}}.$ Then the GC (\ref{gc}) have limit as $\lambda \rightarrow
\lambda _{0}$ of the form
\begin{equation*}
\varepsilon T_{v}^{\prime }\frac{d}{dt}\varsigma ^{(v)}(t)+2i\varsigma
^{(v)}(t)=0,\text{ \ \ \ }t=0,
\end{equation*}
where $T_{v}^{\prime }=\frac{d}{dz}T_{v}.$ The GC also have limit when $%
\varepsilon \rightarrow 0,$ $\lambda \rightarrow \lambda _{0\text{\ }}$%
independently$.$ This limit is the Dirichlet condition $\varsigma
^{(v)}(0)=0.$
\end{theorem}

A simple version of the results presented in this paper (for models
admitting the separation of variables) was published in our paper \cite{MV}.

The next section contains the proofs of Theorem \ref{t1} and \ref{t3}. The
statements of these theorems mostly concern problems with a fixed value of $%
\varepsilon .$ Without loss of the generality, one can assume that $%
\varepsilon =1$ there. The last section is devoted to the proof of Theorem
\ref{t2} on asymptotic behavior of the scattering solutions as $\varepsilon
\rightarrow 0.$ Here the dependence of all objects on $\varepsilon $ is
essential. At the end of the last section, one can find a proof and a short
discussion of Theorem \ref{tlas}.

\section{Analytic properties of the resolvent $R_{\protect\lambda }.$}

We denote by $\Omega _{\varepsilon }^{(a)}$ the following bounded part of $%
\Omega _{\varepsilon }:$%
\begin{equation}
\Omega _{\varepsilon }^{(a)}=\Omega _{\varepsilon }\backslash \underset{%
j\leq m}{\cup }(C_{j,\varepsilon }\cap \{t>a\}).  \label{oma}
\end{equation}
The next lemma will be needed later.

\begin{lemma}
\label{la1}If the homogeneous problem (\ref{a1}), (\ref{a2}) with a real $%
\lambda >0$ has a non-trivial solution $u,$ then either $\sqrt{\lambda }$ is
an eigenvalue of $-\varepsilon ^{2}\Delta $ and $u$ decays exponentially at
infinity, or $\lambda \in \{\lambda _{j,n}\}$ and (\ref{inf}) holds.
\end{lemma}

\textbf{Proof}. $\ $From the Green formula for $u$ and $\overline{u}$ in the
domain $\Omega _{\varepsilon }^{(a)},$ $a>0,$ it follows that
\begin{equation*}
\text{Im}\int_{\partial \Omega _{\varepsilon }^{(a)}}\frac{\partial u}{%
\partial \nu }\overline{u}dS=0,
\end{equation*}
where $\nu $ is the unit normal to $\partial \Omega _{\varepsilon }^{(a)}$
and $dS$ is an element of the surface area. Using the boundary condition (%
\ref{a1}) we arrive at
\begin{equation}
\text{Im}\int_{\partial \Omega _{\varepsilon }^{(a)}\backslash \partial
\Omega _{\varepsilon }}u_{t}\overline{u}dy=0.  \label{gr1}
\end{equation}
This, (\ref{a2}), and the orthogonality of the functions $\varphi _{j,n}$
imply, for $a\rightarrow \infty ,$
\begin{equation*}
\sum_{j,n:\lambda _{j,n}<\lambda }\sqrt{\lambda -\lambda _{j,n}}%
|a_{j,n}|^{2}+O(e^{-\gamma a})=0,
\end{equation*}
which justifies the lemma after taking the limit as $a\rightarrow \infty .$
This completes the proof.

Let $C_{j,\varepsilon }^{\prime }$ be the channel $C_{j,\varepsilon }$
extended along the whole $t$ axis$,$
\begin{equation*}
C_{j,\varepsilon }^{\prime }=\{(t,\varepsilon y):t\in R,\text{ }y\in \omega
_{j}\subset R^{n-1}\}
\end{equation*}
We denote by $R_{\lambda }^{(j)}$ the resolvent (\ref{res}) of the operator $%
-\varepsilon ^{2}\Delta $ in the extended channel $C_{j,\varepsilon
}^{\prime }.$ Let $L_{a}^{2}(C_{j,\varepsilon }^{\prime })$ be the set of
functions from $L^{2}(C_{j,\varepsilon }^{\prime })$ with the support in the
region $|t|\leq a,$ and let $H^{2}(C_{j,\varepsilon }^{b})$ be the Sobolev
space of functions in the domain $C_{j,\varepsilon }^{\prime }\cap
\{b<|t|<b+1\}.$ Consider the operator
\begin{equation}
R_{\lambda }^{(j)}:L_{com}^{2}(C_{j,\varepsilon }^{\prime })\rightarrow
L_{loc}^{2}(C_{j,\varepsilon }^{\prime }).  \label{rl0}
\end{equation}
The following lemma can be easily proved using the method of separation of
variables.

\begin{lemma}
\label{lb1} (1) The operator (\ref{rl0}) admits an analytic continuation
from the upper half plane Im$\lambda >0$ onto the real axis with the branch
points at $\lambda =\lambda _{j,n},$ $n=0,1,...$ .

(2) If $\lambda _{j,m_{j}}<\lambda <\lambda _{j,m_{j}+1}$ and $h\in
L_{com}^{2}(C_{j}^{\prime })$ then\ $R_{\lambda }^{(j)}h$ has the following
behavior as $t\rightarrow \pm \infty $%
\begin{equation}
R_{\lambda }^{(j)}h=\sum_{n=1}^{m_{j}}c_{j,n}^{\pm }e^{i\frac{\sqrt{\lambda
-\lambda _{j,n}}}{\varepsilon }|t|}\varphi _{j,n}(y/\varepsilon
)+O(e^{-\gamma (\varepsilon )|t|}),\text{ \ }\gamma >0,  \label{asc}
\end{equation}
where
\begin{equation}
c_{j,n}^{\pm }=c_{j,n}^{\pm }(h)=\frac{\varepsilon ^{-d}}{2i\sqrt{\lambda
-\lambda _{j,n}}}\int_{\omega _{j,\varepsilon }}\int_{-\infty }^{\infty
}e^{\mp i\frac{\sqrt{\lambda -\lambda _{j,n}}}{\varepsilon }\tau }\varphi
_{j,n}(y/\varepsilon )h(\tau ,y)d\tau dy.  \label{cs}
\end{equation}
(3) Let $\lambda \in l,$ where $l$ is a bounded closed interval of the real
axis such that $\lambda _{j,m_{j}}<\lambda <\lambda _{j,m_{j}+1}$ for all $%
\lambda \in l.$ Let $h\in L_{3\varepsilon }^{2}(C_{j,\varepsilon }^{\prime })
$ and $b\geq 0.$ Then there exist positive constants $c=c(l)$\ and $\gamma
=\gamma (l)$\ which are independent of $\lambda \in l,$ $\varepsilon $ and $%
h,$ and such that the remainder term $r$ in the right-hand side of (\ref{asc}%
) has the estimate
\begin{equation*}
||r||_{H^{2}(C_{j,\varepsilon }^{b})}\leq ce^{-\gamma b/\varepsilon
}||h||_{L_{3\varepsilon }^{2}(C_{j,\varepsilon }^{\prime })}.
\end{equation*}
\end{lemma}

\textbf{Proof of Theorem \ref{t1}.} The statements of the theorem mostly
concern the problem with a fixed value of $\varepsilon .$ Without loss of
generality, we can assume that $\varepsilon =1,$ and we omit $\varepsilon $
in the notations of all objects ($\Omega _{\varepsilon },$ $C_{j,\varepsilon
},$ and so on). The dependence on $\varepsilon $ will be restored in some
parts of the proof, when this dependence on $\varepsilon $ is essential.

\textit{Step 1. Construction of the resolvent.} Let us introduce the
following partition of unity on $\Omega $%
\begin{equation}
\sum_{j=0}^{m}\phi _{j}=1.  \label{a6}
\end{equation}
We fix arbitrary functions $\phi _{j}\in C^{\infty }(\Omega ),$ $1\leq j\leq
m$, such that $\phi _{j}=1$ in the (infinite) channel $C_{j}$ for $t\geq 2,$
$\phi _{j}=0$ in $C_{j}$ for $t\leq 1$ and outside of $C_{j}.$ The function $%
\phi _{0}$ is defined as follows $\phi _{0}=1-\sum_{j\leq m}\phi _{j}.$ We
also need functions $\psi _{j}$ that are equal to one on the supports of $%
\varphi _{j}$, which will allow us to smoothly \ extend functions defined
only on infinite channels or only in a bounded part of $\Omega $ onto the
whole domain $\Omega .$ We fix functions $\psi _{j}\in C^{\infty }(\Omega ),$
$1\leq j\leq m$, such that $\psi _{j}=1$ in the infinite channel $C_{j}$ for
$t\geq 1$ (i. e. on the support of $\phi _{j}$), $\psi _{j}=0$ outside of $%
C_{j}$. Let $\psi _{0}\in C^{\infty }(\Omega )$ be a function\ such that $%
\psi _{0}=1$ on the support of $\phi _{0},$ and $\psi _{0}=0$ in all
infinite channels $C_{j}$ when $t\geq 3.$ Note that
\begin{equation}
\psi _{j}\phi _{j}=\phi _{j},\text{ \ \ }0\leq j\leq m.  \label{a3}
\end{equation}

We construct the parametrix (almost resolvent) for the problem (\ref{a1}) in
the form
\begin{equation}
P_{\lambda }:L^{2}(\Omega )\rightarrow L^{2}(\Omega ),\text{ \ \ \ }%
P_{\lambda }f=\psi _{0}R_{\lambda ^{\prime }}(\phi _{0}f)+\sum_{j=1}^{m}\psi
_{j}R_{\lambda }^{(j)}(\phi _{j}f).  \label{a4}
\end{equation}
where $R_{\lambda ^{\prime }}$ is the resolvent (\ref{res}) of the operator
in $\Omega $ with a fixed $\lambda ^{\prime }=i\sigma ,$ $\sigma >0,$ which
will be chosen later, and $R_{\lambda }^{(j)}$ are resolvents of the
negative Dirichlet Laplacians in $C_{j}$. If $f\in L^{2}(\Omega )$ then $%
\phi _{j}f=0$ outside $C_{j},$ and we consider $\phi _{j}f$ as an element of
$L^{2}(C_{j}).$ Then the operator $R_{\lambda }^{(j)}$ can be applied to $%
\phi _{j}f$ and $R_{\lambda }^{(j)}(\phi _{j}f)\in L^{2}(C_{j}).$ Since $%
\psi _{j}=0$ at the bottom of $C_{j}$ and outside of $C_{j},$ we consider $%
\psi _{j}R_{\lambda }^{(j)}(\phi _{j}f)$ as an element of $L^{2}(\Omega )$
that is equal to zero outside of $C_{j}.$ In this way, the operator $%
P_{\lambda }$ is well defined for $\lambda \notin \lbrack 0,\infty ).$

Let us look for a solution $u\in L^{2}(\Omega )$ of the problem (\ref{a1})
with $\lambda \notin \lbrack 0,\infty )$ in the form of $u=P_{\lambda }h$
with unknown $h\in L^{2}(\Omega ).$ Obviously, $u$ satisfies the Dirichlet
boundary condition since each term in (\ref{a4}), applied to any $h$,
satisfies the Dirichlet boundary condition. The substitution of $P_{\lambda
}h$ for $u$ in equation (\ref{a1}) with $\lambda \notin \lbrack 0,\infty )$
(and $\varepsilon =1$) leads to
\begin{equation*}
(-\Delta -\lambda )P_{\lambda }h=-(\Delta \psi _{0})[R_{\lambda ^{\prime
}}(\phi _{0}h)]-2\nabla \psi _{0}\cdot \nabla \lbrack R_{\lambda ^{\prime
}}(\phi _{0}h)]
\end{equation*}
\begin{equation*}
-\psi _{0}(\Delta +\lambda ^{\prime })[R_{\lambda ^{\prime }}(\phi
_{0}h)]-(\lambda -\lambda ^{\prime })\psi _{0}[R_{\lambda ^{\prime }}(\phi
_{0}h)]
\end{equation*}
\begin{equation*}
-\sum_{j=1}^{m}\left\{ (\Delta \psi _{j})[R_{\lambda }^{(j)}(\phi
_{j}f)]+2\nabla \psi _{j}\cdot \nabla R_{\lambda }^{(j)}(\phi _{j}h)+\psi
_{j}(\Delta +\lambda )[R_{\lambda }^{(j)}(\phi _{j}h)]\right\} =f,
\end{equation*}
Using (\ref{a3}), (\ref{a6}), the last relation can be rewritten in the form
\begin{equation}
h+F_{\lambda }h=f,  \label{a10}
\end{equation}
where
\begin{eqnarray}
F_{\lambda }h &=&-[(\Delta +\lambda -\lambda ^{\prime })\psi
_{0}][R_{\lambda ^{\prime }}(\phi _{0}h)]-2\nabla \psi _{0}\cdot \nabla
\lbrack R_{\lambda ^{\prime }}(\phi _{0}h)]  \notag \\
&&-\sum_{j=1}^{m}\left\{ (\Delta \psi _{j})[R_{\lambda }^{(j)}(\phi
_{j}h)]+2\nabla \psi _{j}\cdot \nabla R_{\lambda }^{(j)}(\phi _{j}h)\right\}
.  \label{a7}
\end{eqnarray}

Let us show that the operator
\begin{equation}
F_{\lambda }:L^{2}(\Omega )\rightarrow L^{2}(\Omega ),\text{ \ \ \ }\lambda
\notin \lbrack \lambda _{0},\infty ),  \label{a9}
\end{equation}
is compact and depends analytically on $\lambda $. Indeed, the resolvents $%
R_{\lambda ^{\prime }}$ and $R_{\lambda }^{(j)}$ map any function $f\in
L^{2} $ into the solution of the problem (\ref{a1}) in the domains $\Omega ,$
$C_{j},$ respectively. Thus, these operators are bounded as operators from $%
L^{2}$ into the Sobolev spaces $H^{2}.$ Since the formula (\ref{a7})
contains at most first derivatives of the resolvents, the operator $%
F_{\lambda },$\ \ $\lambda \notin \lbrack \lambda _{0},\infty ),$ is bounded
if it is considered as an operator from $L^{2}(\Omega )$ into the Sobolev
space $H^{1}(\Omega )$. Since $\nabla \psi _{0}=\nabla \psi _{j}=0$ at
points $x\in C_{j}$ with $t>3,$ from (\ref{a7}) it follows that, for any
infinite channel $C_{j},$
\begin{equation}
F_{\lambda }h=0,\text{ \ \ }x\in C_{j}\cap \{t>3\}.  \label{a11}
\end{equation}
Hence, the Sobolev imbedding theorem implies that the operator (\ref{a9}) is
compact. The analyticity of the operator (\ref{a9}) is obvious since the
operators $R_{\lambda }^{(j)}$ depend analytically on $\lambda ,$ and $%
R_{\lambda ^{\prime }}$ does not depend on $\lambda .$

Now we put $\lambda =\lambda ^{\prime }=i\sigma $ and show that $%
||F_{i\sigma }||\rightarrow 0$ as $\sigma \rightarrow \infty .$ In fact,
since the norm of the resolvent does not exceed the inverse distance from
the spectrum, we have that
\begin{equation}
||R_{\lambda ^{\prime }}||,\text{ }||R_{\lambda ^{\prime }}^{(j)}||\leq
1/\sigma ,  \label{es1}
\end{equation}
where the first norm is considered in the space $L^{2}(\Omega )$ and the
second one is in the space $L^{2}(C_{j}).$ Multiplying the equation (\ref{a1}%
), considered in the domain $\Omega $ or $C_{j},$ by $u$ and integrating
over the domain, we get the following relation for the functions $%
u=R_{\lambda ^{\prime }}f$ and $u=R_{\lambda ^{\prime }}^{(j)}f,$
respectively:
\begin{equation*}
||\nabla u||_{L^{2}}^{2}-i\sigma \text{ }||u||_{L^{2}}^{2}=\int ufdx,
\end{equation*}
which implies that
\begin{equation*}
||\nabla u||_{L^{2}}^{2}\leq |\int ufdx|\leq ||u||_{L^{2}}||f||_{L^{2}}.
\end{equation*}
Thus,
\begin{equation}
||R_{\lambda ^{\prime }}f||_{H^{1}(\Omega )},\text{ }||R_{\lambda ^{\prime
}}^{(j)}f||_{H^{1}(C_{j})}\leq C\sigma ^{-1/2}||f||_{L^{2}}.  \label{es2}
\end{equation}
Since the formula (\ref{a7}) contains at most first derivatives of the
resolvents, estimates (\ref{es1}), (\ref{es2}) imply that $||F_{i\sigma
}||\rightarrow 0$ as $\sigma \rightarrow \infty .$

We fix $\lambda ^{\prime }=i\sigma $ in (\ref{a4}) in such a way that $%
||F_{\lambda ^{\prime }}||<1.$ Then from the analytic Fredholm theorem it
follows that the operator
\begin{equation}
(E+F_{\lambda })^{-1}:L^{2}(\Omega )\rightarrow L^{2}(\Omega ),\text{ \ \ }%
\lambda \notin \lbrack \lambda _{0},\infty ),  \label{a13}
\end{equation}
exists and depends meromorphically on $\lambda .$ From here, (\ref{a4}) and (%
\ref{a10}) the following representation for the resolvent follows
\begin{equation}
R_{\lambda }=P_{\lambda }(E+F_{\lambda })^{-1},\ \ \lambda \notin \lbrack
\lambda _{0},\infty ).  \label{a12}
\end{equation}

\textit{Step 2. Analytic continuation of the resolvent.} In order to extend
the operator (\ref{b2}) meromorphically into the lower half plane Im$k<0$ we
need to repeat the arguments used to justify (\ref{a12}). Consider the space
$L_{a}^{2}(\Omega )$ of functions $f\in L^{2}(\Omega )$ with supports in $%
\Omega ^{(a)}$ (see (\ref{oma})), i.e. $f=0$ in the infinite channels $C_{j}$
when $t>a.$ Let $f\in L_{a}^{2}(\Omega ).$ Without loss of generality, one
can assume that $a>3.$ \ Then (\ref{a10}) and (\ref{a11}) imply that $h\ $is
also supported in $\Omega ^{(a)},$ i.e. $F_{\lambda }$ can be considered as
an operator in $L_{a}^{2}(\Omega ):$%
\begin{equation*}
F_{\lambda }:L_{a}^{2}(\Omega )\rightarrow L_{a}^{2}(\Omega ),\text{ \ \ }%
\lambda \notin \lbrack 0,\infty ).
\end{equation*}
Let $\chi =\chi _{a}(t)$ be a function equal to one when $t\leq a$ and zero
when $t>a.$ From Lemma \ref{lb1} it follows that the operators
\begin{equation*}
\chi R_{k^{2}}^{(j)}:L_{a}^{2}(C_{j})\rightarrow L_{a}^{2}(C_{j}),\text{ \ \
Im}k>0,
\end{equation*}
admit an analytic continuation into the lower half plane with the branch
points at $k=\pm \sqrt{\lambda _{j,n}}$. Further, $u=R_{k^{2}}^{(j)}f$
satisfies equation (\ref{a1}) with $\lambda =k^{2}$ for all complex $k\in
\mathbf{C},$ and therefore the operators
\begin{equation*}
\chi R_{k^{2}}^{(j)},\text{ }\chi \nabla
R_{k^{2}}^{(j)}:L_{a}^{2}(C_{j})\rightarrow L_{a}^{2}(C_{j}),\text{ \ \ \ }%
k\in \mathbf{C,}
\end{equation*}
are compact and analytic in the complex plane $\mathbf{C}.$ Since $\chi =1$
on the supports of $\nabla \psi _{j},$ $0\leq j\leq m,$ we can insert the
factor $\chi $ on the left of all the resolvents $R_{\lambda }^{(j)}$ in (%
\ref{a7}). From here it follows that the operator
\begin{equation*}
F_{k^{2}}:L_{a}^{2}(\Omega )\rightarrow L_{a}^{2}(\Omega ),\text{ \ \ \ }%
k\in \mathbf{C,}
\end{equation*}
is compact and analytic with branch points at \ $k=\pm \sqrt{\lambda _{j,n}}%
. $ Hence, the operator
\begin{equation}
(E+F_{k^{2}})^{-1}:L_{a}^{2}(\Omega )\rightarrow L_{a}^{2}(\Omega ),\text{ \
\ \ }k\in \mathbf{C},  \label{a14}
\end{equation}
is meromorphic with the branch points at $k=\pm \sqrt{\lambda _{j,n}}.$ \
Together with (\ref{a4}), (\ref{a12}) and the analyticity of the operators $%
R_{k^{2}}^{(j)}:L_{a}^{2}(C_{j})\rightarrow L_{loc}^{2}(C_{j}),$ $k\in
\mathbf{C},$ this implies that the operator (\ref{b2}) admits a meromorphic
continuation to the lower half plane with the branch points at $k=\pm \sqrt{%
\lambda _{j,n}}$ and poles determined by the poles of the operator (\ref{a14}%
). Obviously, the poles of the operator (\ref{a14}) may have a limiting
point only at $\lambda =\infty .$

\textit{Step 3. Spectral analysis. }First of all note that the existence of
the meromorphic extension of the operator (\ref{b2}) together with the Stone
formula immediately imply that the operator $H=-\Delta $ does not have
singular spectrum. The proof of this fact can be found in \cite{RID} (see
Theorem XIII.20).

In order to prove the part of statement (1) of the theorem concerning the
absolutely continuous spectrum of the operator $H=-\Delta $, we split the
domain $\Omega $ into pieces by introducing cuts along the bases $t=0$ of
all infinite channels. We denote the new (not connected) domain by $\Omega
^{\prime },$ and denote the negative Dirichlet Laplacian in $\Omega ^{\prime
}$ by $H^{\prime },$ i. e. $H^{\prime }$ is obtained from $H$ by introducing
additional Dirichlet boundary conditions on the cuts. Obviously, the
operator $H$ has the absolutely continuous spectrum described in statement
(1) of the theorem. Thus, it remains to show that the wave operators for the
couple $H,$ $H^{\prime }$ exist and complete. The justification of the
existence and completeness of the wave operators can be found in \cite{BIR}.
Another option is to derive the latter fact\ independently using Birman
theorem stating that the validity of the inclusion 
\begin{equation}
(H-\lambda )^{-n}-(H^{\prime }-\lambda )^{-n}\in J_{1}  \label{t}
\end{equation}
for some $\lambda $ and $n\geq 1$ implies the existence and completeness of
the wave operators$.$ Here $J_{1}$ is the space of operators of the trace
class. The inclusion (\ref{t}) can be derived from (\ref{a12}) and a similar
formula for the resolvent of the operator $H^{\prime }.$ This completes the
proof of the statement about the absolutely continuous spectrum.

The discreteness of the set $\{\lambda ^{j,\varepsilon }\}$ of eigenvalues
follows from the fact that the operator (\ref{b2}) is meromorphic in $%
\lambda $ and has poles at $\{\lambda ^{j,\varepsilon }\}.$ The existence of
the poles at $\{\lambda ^{j,\varepsilon }\}$ can be derived from the Stone
formula. Another proof will be given below.

Let us prove the part of statement (1) concerning the spider domains. If $%
\Omega _{\varepsilon }$ is a spider domain, then there exists a point $%
\widehat{x}(\varepsilon )$ and an $\varepsilon $-independent domain $\Omega $
such that the transformation (see~(\ref{hom})) \ 
\begin{equation}
L_{\varepsilon }:x\rightarrow \widehat{x}(\varepsilon )+\varepsilon x,
\label{te}
\end{equation}
maps $\Omega $ into $\Omega _{\varepsilon }.$ In order to stress the fact
that the operator $H=-\varepsilon ^{2}\Delta $ in the domain $\Omega
_{\varepsilon }$ depends on $\varepsilon ,$ we shall denote it by $%
H^{(\varepsilon )}.$ The operator $-\Delta $ in the domain $\Omega $ shall
be denoted by $H^{(1)}$. Obviously,
\begin{equation}
H^{(\varepsilon )}=L_{\varepsilon }H^{(1)}L_{\varepsilon }^{-1},
\label{hom1}
\end{equation}
and this implies the independence of the eigenvalues of the operator $%
H^{(\varepsilon )}$ of $\varepsilon .$ This completes the proof of statement
(1).

\textit{Step 4, real poles of the resolvent. }The first part of statement
(2) about the existence of the analytic extension of the resolvent was
justified in step 2 of the proof. Now we are going to prove the second part
of that statement concerning the set of real poles of the operator (\ref{b2}%
). We denote this set of poles by $K.$ Let us assume that either $u$ is an
eigenfunction of the operator $H=-\Delta $ with an eigenvalue $\lambda
=\lambda ^{\prime }>0$ or $u$ is a non-trivial solution of the homogeneous
problem (\ref{a1}), (\ref{inf}) with $\lambda =\lambda ^{\prime }>0$ (recall
that we assume that $\varepsilon =1).$ We are going to show that $k=\pm
\sqrt{\lambda ^{\prime }}\in K.$ Consider the restrictions $u_{j}$ of $u$ to
the cylinders $C_{j},$ $1\leq j\leq m.$ Let $v_{j}\in L^{2}(C_{j})$ be the
solution of the problem
\begin{equation*}
(-\Delta -\lambda )v_{j}=0,\text{ \ \ \ \ }x\in C_{j};\text{ \ \ \ \ }v_{j}=0%
\text{ \ on }\partial ^{\prime }C_{j},\text{ \ \ }v_{j}=u_{j}\text{ \ when }%
t=0,
\end{equation*}
where\ $\lambda \notin \lbrack 0,\infty ),$ and$\ \partial ^{\prime }C_{j}$
is the lateral boundary of $C_{j}.$ The solution $v_{j}\in L^{2}(C_{j})$ of
this problem is unique and can be found by separation of variables. The
function $u_{j}$ satisfies the same equation with the fixed $\lambda
=\lambda ^{\prime }$ and the same boundary conditions. It is also defined
uniquely by its values at $t=0$ and can be found by separation of variables.
This implies that $v_{j}$ converges to $u$ as $\lambda \rightarrow \lambda
^{\prime }+i0.$ Since $u$ is a solution of a homogeneous elliptic problem, $%
u\in C^{\infty }.$ Thus, $u_{j}\ $is infinitely smooth when $t=0,$ and the
convergence $v_{j}\rightarrow u_{j}$ takes place, for example, in the
Sobolev space $H^{2}$ on the part of the cylinder $C_{j}$ where $0\leq t\leq
2.$ Let
\begin{equation*}
v=\sum_{j=1}^{m}\phi _{j}v_{j}+\phi _{0}u\in L^{2}(C_{j}),\ \ \lambda \notin
\lbrack 0,\infty ),
\end{equation*}
where $\{\phi _{j}\}$ is the partition of unity which was introduced above.
The function $u$ can not be equal to zero identically on $\Omega \backslash
\cup C_{j}$ due to the uniqueness of the solution of the Cauchy problem for
the operator $-\Delta -\lambda ^{\prime }.$ Thus
\begin{equation}
||v||_{L^{2}(\Omega \backslash \cup C_{j})}=||u||_{L^{2}(\Omega \backslash
\cup C_{j})}=c_{0}>0.  \label{b4}
\end{equation}
On the other hand,
\begin{equation*}
(-\Delta -\lambda )v=-\sum_{j=1}^{m}[(\Delta \phi _{j})v_{j}+2\nabla \phi
_{j}\cdot \nabla v_{j}]-(\lambda -\lambda ^{\prime })\phi _{0}u-(\Delta \phi
_{0})u-2\nabla \phi _{0}\cdot \nabla u.
\end{equation*}
Thus, $(-\Delta -\lambda )v\in L_{a}^{2}(\Omega ).$ From the convergence $%
v_{j}\rightarrow u$ and (\ref{a6}) it follows that $(-\Delta -\lambda )v$
tends to zero in $L_{a}^{2}(\Omega )$ as $\lambda \rightarrow \lambda
^{\prime }+i0.$ Together with (\ref{b4}) this provides the existence of the
pole of the operator (\ref{b2}) at $k=\sqrt{\lambda ^{\prime }}.$ The pole
at $k=-\sqrt{\lambda ^{\prime }}$ exists due to the relation $R_{\lambda }=%
\overline{R_{\overline{\lambda }}}.$ Hence, $\pm \sqrt{\lambda ^{\prime }}%
\in K.$

Now let us assume that at least one of the points $\pm \sqrt{\lambda
^{\prime }}$ belongs to $K.$ The relation $R_{\lambda }=\overline{R_{%
\overline{\lambda }}}$ implies that the second point also belongs to $K,$
i.e. $\sqrt{\lambda ^{\prime }}\in K,\ $and there exist $a>0$ $\ $and $f\in
L_{a}^{2}(\Omega )$ such that
\begin{equation}
w;=R_{\lambda }f=\frac{u(x)}{(\lambda -\lambda ^{\prime })^{n}}+\frac{%
v(x,\lambda )}{(\lambda -\lambda ^{\prime })^{n-1}};\text{ \ \ }%
||v||_{L^{2}(\Omega ^{(a+2)})}\leq c,\text{ \ \ \ }\lambda \rightarrow
\lambda ^{\prime }+i0,  \label{b5}
\end{equation}
where $n\geq 1$ and $u$ does not vanish identically. In fact, $n$ can not
exceed one, but it is not important for us now. Obviously,
\begin{equation}
(-\Delta -\lambda ^{\prime })u=0,\text{ \ }x\in \Omega ;\text{ \ \ \ }u=0%
\text{ \ on }\partial \Omega .  \label{ab5}
\end{equation}
From here and Lemma \ref{la1} it follows that in order to complete the proof
of the second statement of the theorem it is sufficient to show that the
asymptotic expansion (\ref{a2}) holds for the function $u.$

Note that (\ref{ab5}) implies that $u\in C^{\infty }.$ Since $f=0$ in all
infinite channels $C_{j}$ when $t>a,$ from relation (\ref{b5}) it follows
that
\begin{equation*}
(-\Delta -\lambda )v=(\lambda -\lambda ^{\prime })u,\text{ \ \ }x\in
C_{j}\cap \{t>a\};\text{ \ \ \ }v=0\text{ \ \ on }\partial \Omega .
\end{equation*}
From here, the estimate in (\ref{b5}), and standard local a priory estimates
for solutions of elliptic problems it follows that for any vector $\alpha $%
\begin{equation*}
|\frac{\partial ^{\alpha }v}{\partial x^{\alpha }}|\leq c(\alpha ),\text{ \
\ }x\in C_{j}\cap \{a+\frac{3}{2}>t>a+\frac{1}{2}\},\text{ \ \ }\lambda
\rightarrow \lambda ^{\prime }+i0,
\end{equation*}
and therefore
\begin{equation}
\frac{\partial ^{\alpha }[(\lambda -\lambda ^{\prime })^{n}w]}{\partial
x^{\alpha }}\rightarrow \frac{\partial ^{\alpha }u}{\partial x^{\alpha }}
\label{lll}
\end{equation}
uniformly on $C_{j}\cap \{t=a+1\}$ as $\lambda \rightarrow \lambda ^{\prime
}+i0.$ We restrict the functions $(\lambda -\lambda ^{\prime })^{n}w$ and $u$
to $C_{j}\cap \{t=a+1\}$ and expand the restrictions with respect to the
basis $\{\varphi _{j,n}\}$ of the operator $-\Delta $ in the cross section
of the channel $C_{j}.$ Let $\gamma _{j,n}(\lambda )$ and $\gamma _{j,n}^{0}$
be the coeficients of these expansions. Then (\ref{lll}) implies that for
any $\beta ,$
\begin{equation}
|\gamma _{j,n}(\lambda )-\gamma _{j,n}^{0}|<c_{\beta }n^{-\beta },\text{ \ \
}\lambda \rightarrow \lambda ^{\prime }+i0.  \label{b8}
\end{equation}

The function $\widehat{w}:=(\lambda -\lambda ^{\prime })^{n}w$ satisfies the
following relations in $C_{j}\cap \{t\geq a+1\}:$
\begin{equation*}
(-\Delta -\lambda )\widehat{w}=0,\text{\ \ \ }\widehat{w}=0\text{ \ for\ \ }%
x\in \partial C_{j}\cap \{t>a+1\},\text{ \ \ }\widehat{w}|_{t=a+1}=\sum_{n}%
\gamma _{j,n}(\lambda )\varphi _{j,n}(y),
\end{equation*}
where $\lambda \notin \lbrack 0,\infty ).$ One can find the solution $%
\widehat{w}\in L^{2}$ of this problem by the method of separation of
variables and then pass to the limit as $\lambda \rightarrow \lambda
^{\prime }+i0$ using (\ref{b8}). This leads to the asymptotic expansion (\ref
{a2}) for $u$ and completes the proof of the second statement of Theorem \ref
{t1}.

\textit{Step 5, the proof of the last two statements of the theorem.} If $k=%
\sqrt{\lambda ^{\prime }},$ $\lambda ^{\prime }>0,$ is not a pole or a
branch point of $R_{k^{2}}$ then
\begin{equation*}
w:=R_{\lambda }f=u(x)+(\lambda -\lambda ^{\prime })v(x,\lambda );\text{ \ \ }%
||v||_{L^{2}(\Omega ^{(a)})}\leq c(a),\text{ \ \ \ }\lambda \rightarrow
\lambda ^{\prime }+i0,
\end{equation*}
where $a>0$ is arbitrary and
\begin{equation*}
(-\Delta -\lambda ^{\prime })u=f,\text{ \ }x\in \Omega ;\text{ \ \ \ }u=0%
\text{ \ on }\partial \Omega .
\end{equation*}
In order to prove the third statement of the theorem, we need only to show
that the asymptotic expansion (\ref{a2}) holds for $u.$ It can be done
exactly in the same way as it was done for function $u$ in (\ref{b5}) by
representing $u$ in $C_{j}\cap \{t>a+1\}$ as the limit of functions $w$ as \
$\lambda \rightarrow \lambda ^{\prime }+i0.$

In order to prove the last statement of the theorem one can look for the
solution $\Psi =\Psi _{s,k}$ of the scattering problem in the form
\begin{equation*}
\Psi =\phi _{s}e^{-i\sqrt{\lambda -\lambda _{s,k}}t}\varphi _{s,k}(y)+u,
\end{equation*}
where $\phi _{s}$ is the function from the partition of unity (\ref{a6}).
This reduces problem (\ref{b9}), (\ref{b10}) to the uniquely solvable
problem (\ref{a1}), (\ref{a2}) for $u$.

This completes the proof of Theorem \ref{t1}.

\begin{proposition}
\label{p1}Let $\Psi _{s,k}$ and $\Psi _{s^{\prime },k^{\prime }}$ be two
scattering solutions, and let $a_{j,n}^{s,k}$ be the transmission
coefficients for the scattering solution $\Psi _{s,k}$. Then

1) The following energy conservation law is valid:
\begin{equation*}
\sum_{j,n}\sqrt{\lambda -\lambda _{j,n}}|a_{j,n}^{s,k}|^{2}=\sum_{j=1}^{m}%
\sum_{n=0}^{m_{j}}\sqrt{\lambda -\lambda _{j,n}}|a_{j,n}^{s,k}|^{2}=\sqrt{%
\lambda -\lambda _{s,k}}.
\end{equation*}

2) If these solutions correspond to different incident waves ($(s,k)\neq
(s^{\prime },k^{\prime })$), then
\begin{equation*}
\sum_{j,n}\sqrt{\lambda -\lambda _{j,n}}a_{j,n}^{s,k}\overline{a}%
_{j,n}^{s^{\prime },k^{\prime }}=0.
\end{equation*}
\end{proposition}

\textbf{Proof.} Since the statement concerns the problem with a fixed value
of $\varepsilon ,$ one can put $\varepsilon =1$ and omit $\varepsilon $ in
the notations $\Omega _{\varepsilon },$ $\Omega _{\varepsilon }^{(a)}$.
Green's formula for $\Psi _{s,k}$ and $\overline{\Psi }_{s^{\prime
},k^{\prime }}$ in the domain $\Omega ^{(a)}$ implies, similarly to (\ref
{gr1}), that
\begin{equation*}
\int_{\partial \Omega ^{(a)}\backslash \partial \Omega }[(\Psi _{s,k})_{t}%
\overline{\Psi }_{s^{\prime },k^{\prime }}-\Psi _{s,k}(\overline{\Psi }%
_{s^{\prime },k^{\prime }})_{t}]dy=0.
\end{equation*}
From here, (\ref{b10}), and the orthogonality of the functions $\varphi
_{j,n}$ it follows that
\begin{equation*}
\sum_{j,n}\sqrt{\lambda -\lambda _{j,n}}a_{j,n}^{s,k}\overline{a}%
_{j,n}^{s^{\prime },k^{\prime }}-\sqrt{\lambda -\lambda _{s,k}}\overline{a}%
_{j,n}^{s,k}e^{-2i\sqrt{\lambda -\lambda _{s,k}}a}
\end{equation*}
\begin{equation*}
+\sqrt{\lambda -\lambda _{s^{\prime },k^{\prime }}}a_{j,n}^{s^{\prime
},k^{\prime }}e^{2i\sqrt{\lambda -\lambda _{s^{\prime },k^{\prime }}}a}-%
\sqrt{\lambda -\lambda _{s,k}}\delta +O(e^{-\gamma a})=0,\text{ \ \ \ }%
a\rightarrow \infty ,
\end{equation*}
where $\delta =1$ if $\left(
\begin{array}{c}
s \\
k
\end{array}
\right) =\left(
\begin{array}{c}
s^{\prime } \\
k^{\prime }
\end{array}
\right) ,$ and $\delta =0$ otherwise. We take the average with respect to $%
a\in (A,2A)$ and pass to the limit as $A\rightarrow \infty .$ Then we get
\begin{equation*}
\sum_{j,n}\sqrt{\lambda -\lambda _{j,n}}a_{j,n}^{s,k}\overline{a}%
_{j,n}^{s^{\prime },k^{\prime }}=\sqrt{\lambda -\lambda _{s,k}}\delta ,
\end{equation*}
which justifies both statements of the proposition. This completes the proof.

\textbf{Proof of Theorem \ref{t3}.} Proposition \ref{p1} is equivalent to
the relation $A^{\ast }A=I$ for $A=D^{1/2}TD^{-1/2},$ which provides the
unitarity of the matrix $A.$ If one applies Green's formula to the
scattering solutions $\Psi _{s,k}$ and $\Psi _{s^{\prime },k^{\prime }},$
then the arguments used in the proof of Proposition \ref{p1} lead to the
symmetry of $D^{1/2}TD^{-1/2}.$ This completes the proof of Theorem \ref{t3}.

\section{Asymptotic behavior of scattering solutions as $\protect\varepsilon %
\rightarrow 0$.}

We start with a study of scattering solutions in spider domains $\Omega
_{\varepsilon }$

\begin{lemma}
\label{l21}Theorem \ref{t2} is valid for spider domains.
\end{lemma}

\textbf{Proof.} The transformation $L_{\varepsilon }^{-1},$\ see (\ref{te}),
maps the spider domain $\Omega _{\varepsilon }$ into the $\varepsilon -$%
independent domain $\Omega $ with the channels $C_{j},$ $1\leq j\leq m,$ The
coordinates $(\widehat{t},\widehat{y})$ in $C_{j}$ are related to
coordinates $(t,y)$ in $C_{j,\varepsilon }$ via the formulas
\begin{equation}
\widehat{t}=t/\varepsilon ,\text{ \ \ }\widehat{y}=y/\varepsilon .
\label{te1}
\end{equation}
The scattering solution $\widehat{\Psi }=\widehat{\Psi }_{s,k}$ of the
problem in $\Omega $ has the form similar to (\ref{b10}):
\begin{eqnarray*}
\widehat{\Psi }_{s,k} &=&\delta _{s,j}e^{-i\sqrt{\lambda -\lambda _{s,k}}%
\widehat{t}}\varphi _{s,k}(\widehat{y})+\sum_{n=0}^{m_{j}}t_{j,n}e^{i\sqrt{%
\lambda -\lambda _{j,n}}\widehat{t}}\varphi _{j,n}(\widehat{y})+O(e^{-\gamma
\widehat{t}}),\text{ \ \ } \\
x &\in &C_{j},\text{ \ \ }\widehat{t}\rightarrow \infty .
\end{eqnarray*}
Since $\widehat{\Psi }_{s,k}$ is a smooth function, the remainder term $%
\widehat{r}$ in the formula above can be estimated for all values of $%
\widehat{t}:$%
\begin{equation*}
|\widehat{r}|\leq Ce^{-\gamma \widehat{t}},\text{ \ \ }x\in C_{j}.
\end{equation*}
Since the scattering solutions in the domains $\Omega _{\varepsilon }$ and $%
\Omega $ are related via the formula $\Psi _{s,k}(x)=\widehat{\Psi }%
_{s,k}(L_{\varepsilon }^{-1}x)$, it follows that
\begin{eqnarray}
\Psi _{s,k}^{(\varepsilon )} &=&\delta _{s,j}e^{-i\frac{\sqrt{\lambda
-\lambda _{s,k}}}{\varepsilon }t}\varphi _{s,k}(y/\varepsilon
)+\sum_{n=0}^{m_{j}}t_{j,n}e^{i\frac{\sqrt{\lambda -\lambda _{j,n}}}{%
\varepsilon }t}\varphi _{j,n}(y/\varepsilon )+r^{(\varepsilon )},\text{ \ \ }
\notag \\
|r^{(\varepsilon )}| &\leq &Ce^{-\gamma t/\varepsilon },\text{ \ \ }x\in
C_{j}.  \label{as3}
\end{eqnarray}
Thus, the asymptotic expansion (\ref{psias}), (\ref{uv}) is valid, and it
only remains to show that the GC (\ref{gc}) holds for vectors $\varsigma
=\varsigma _{s,k}$ determined by (\ref{as3}) (the definition of these
vectors is given in the paragraph above formula (\ref{gc})). We form the
matrix $\Sigma =\Sigma (t)$ with columns $\varsigma _{s,k}$ taking them in
the same order as the order chosen for elements in each of these vectors
(first we put columns with $s=1$ and $k=1,2,...,m_{1},$ then columns with $%
s=2,$ and so on). From (\ref{as3}) it follows that
\begin{equation*}
\Sigma (0)=I+T,\text{ \ \ }\Sigma ^{\prime }(0)=\frac{i}{\varepsilon }%
D(-I+T),
\end{equation*}
where $T$ is the scattering matrix, $I$ is the identity matrix of the same
size, and $D$ is the diagonal matrix of the same size with elements $\sqrt{%
\lambda -\lambda _{j,n}}$ on the diagonal. Hence, $\varepsilon
(I+T)D^{-1}\Sigma ^{\prime }(0)+i(I-T)\Sigma (0)=0$ and GC (\ref{gc}) holds
for the columns of the matrix $\Sigma .$ This completes the proof of the
lemma.

The following two lemmas about spider domains will be needed in order to
prove Theorem \ref{t2} for general domains. Let $R_{\lambda }$ be the
resolvent of the operator $H=-\varepsilon ^{2}\Delta $ in a spider domain $%
\Omega _{\varepsilon },$ and let $\widehat{R}_{\lambda }$ be the resolvent
of the similar operator $H=-\Delta $ in the domain $\Omega $ which is the
image of $\Omega _{\varepsilon }$ under the map $L_{\varepsilon }^{-1},$ see
(\ref{te}). Note that the operator $R_{\lambda }$ and its domain, $%
L^{2}(\Omega _{\varepsilon }),$ depend on $\varepsilon ,$ while the operator
$\widehat{R}_{\lambda }$ is $\varepsilon $-independent. Formula (\ref{hom1})
implies

\begin{lemma}
\label{l22}The following relation holds
\begin{equation*}
R_{\lambda }=L_{\varepsilon }\widehat{R}_{\lambda }L_{\varepsilon }^{-1}.
\end{equation*}
\end{lemma}

Let us fix $m$ constants $t_{j}>0,$ $1\leq j\leq m.$ Let $\Omega
_{\varepsilon }$ be a spider domain with the channels $C_{j,\varepsilon },$ $%
1\leq j\leq m.$ Consider slices $D_{j,\varepsilon }$ of $C_{j,\varepsilon }$
defined by the inequalities $|t-t_{j}|\leq 3\varepsilon $. Let $\Omega
_{\varepsilon }^{\prime }$ be a bounded domain which is obtained from $%
\Omega _{\varepsilon }$ by cutting off the infinite parts of channels $%
C_{j,\varepsilon }$ on which $t\geq \frac{3}{4}t_{j.}$ Let a function $h\in
L^{2}(\Omega _{\varepsilon })$ be supported in one of the domains $%
D_{j,\varepsilon },$ for example, with $j=s.$

Below, when the resolvent $R_{\lambda }$ of the operator $H=-\varepsilon
^{2}\Delta $ in $\Omega _{\varepsilon }$ is considered with $\lambda $
belonging to the continuous spectrum of the operator, $R_{\lambda }$ is
understood in the sense of the analytic extension described in Theorem \ref
{t1}. We denote the Sobolev spaces of functions which are square integrable
together with their derivatives of up to the second order by $H^{2}(\Omega
_{\varepsilon }^{\prime })$ and $H^{2}(D_{j,\varepsilon })$ .

\begin{lemma}
\label{l23} Let $\Omega _{\varepsilon }$ be a spider domain. Let $l$ be a
bounded closed interval of the $\lambda $-axis that does not contain points $%
\lambda _{j,n},$ and let a function $h\in L^{2}(\Omega _{\varepsilon })$ be
supported in the domain $D_{s,\varepsilon }.$ Then there exists $\gamma
=\gamma (\omega _{j},l)>0$ such that
\begin{equation}
(1)\text{ \ \ \ }R_{\lambda }h=\sum_{k=0}^{m_{s}}c_{s.k}\Psi _{s,k}+r_{0}%
\text{ \ in \ \ }\Omega _{\varepsilon }^{\prime },\text{ \ \ }|r_{0}(x)|\leq
\frac{Ce^{-\gamma /\varepsilon }}{\Pi _{\lambda ^{j}\in l}|\lambda -\lambda
^{j}|}||h||_{L^{2}(\Omega _{\varepsilon })},  \label{131}
\end{equation}
where $\Psi _{s,k}$ are scattering solutions, the coefficients $%
c_{s.k}=c_{s.k}^{-}(h)$ are given by (\ref{cs}), and $\lambda ^{j}$ are
eigenvalues of the operator $H$ in $\Omega _{\varepsilon }$ (see statement
(1) of Theorem \ref{t1});
\begin{equation}
(2)\text{ \ \ \ }R_{\lambda }h=\delta _{s,j}R_{\lambda
}^{(s)}h+\sum_{k=0}^{m_{s}}c_{s.k}\sum_{n=0}^{m_{j}}t_{j,n}^{s,k}e^{i\frac{%
\sqrt{\lambda -\lambda _{j,n}}}{\varepsilon }t}\varphi _{j,n}(y/\varepsilon
)+r_{j}\text{ \ in \ \ }D_{j,\varepsilon },\text{ \ \ }  \label{132}
\end{equation}
\begin{equation*}
\text{where \ \ }||r_{j}||_{H^{2}(D_{j,\varepsilon })}\leq \frac{Ce^{-\gamma
/\varepsilon }}{\Pi _{\lambda ^{j}\in l}|\lambda -\lambda ^{j}|}%
||h||_{L^{2}(\Omega _{\varepsilon })}.
\end{equation*}
Here $\delta _{s,j}$ is the Kronecker symbol ($\delta _{s,j}=1$ if $s=j,$ $%
\delta _{s,j}=0$ if $s\neq j)$, $R_{\lambda }^{(s)}$ is the resolvent of $%
-\Delta $ in the extended channel $C_{s,\varepsilon }^{\prime }$ (channel $%
C_{s,\varepsilon }$ extended to $-\infty $ along the $t$ axis), $%
c_{s.k}=c_{s.k}^{-}(h),$ and $t_{j,n}^{s,k}$ are the transmission
coefficients (see the remark following definition \ref{d2}).
\end{lemma}

\textbf{Proof.} Let a function $\alpha \in C^{\infty }(\Omega _{\varepsilon
})$ have the form:\ $\alpha =1$ in $C_{s,\varepsilon }$ when $t>\frac{7}{8}%
t_{j}+\varepsilon ,$ $\alpha =0$ in $\Omega _{\varepsilon }\backslash
C_{s,\varepsilon },$ and $\alpha =0$ in $C_{s,\varepsilon }$ when $t<\frac{7%
}{8}t_{j}.$ Consider the function
\begin{equation}
u=\alpha R_{\lambda }^{(s)}h+\sum_{k=0}^{m_{s}}c_{s.k}[\Psi _{s,k}-\alpha
e^{-i\frac{\sqrt{\lambda -\lambda s,k}}{\varepsilon }t}\varphi
_{s,k}(y/\varepsilon )].  \label{asd}
\end{equation}
Obviously, $u=0$ on $\partial \Omega _{\varepsilon },$ since each term in
the right hand side above satisfies the Dirichlet boundary condition.
Furthermore,
\begin{equation}
-\varepsilon ^{2}\Delta u-\lambda u=h-\varepsilon ^{2}[\nabla \alpha \cdot
\nabla R_{\lambda }^{(s)}h+(\Delta \alpha )R_{\lambda }^{(s)}h-\nabla \alpha
\cdot \nabla g-(\Delta \alpha )g],  \label{as}
\end{equation}
where
\begin{equation*}
g=\sum_{k=0}^{m_{s}}c_{s.k}e^{-i\frac{\sqrt{\lambda -\lambda s,k}}{%
\varepsilon }t}\varphi _{s,k}(y/\varepsilon ).
\end{equation*}
The right hand side in (\ref{as}) has the form $h+h_{1},$ where $h_{1}$ is
supported in the slice $\frac{7}{8}t_{s}\leq t\leq \frac{7}{8}%
t_{s}+\varepsilon $ of $C_{s,\varepsilon }.$ From Lemma \ref{lb1} it follows
that
\begin{equation*}
||h_{1}||_{L^{2}(\Omega _{\varepsilon })}\leq Ce^{-\gamma /\varepsilon
}||h||_{L^{2}(\Omega _{\varepsilon })}.
\end{equation*}
It is also clear that the behavior of the function $u$ at infinity is
described by (\ref{a2}). Hence, $u=$ $R_{\lambda }(h+h_{1})$ due to the
statement (3) of Theorem \ref{t1}. From here and (\ref{asd}) it follows that
\begin{equation}
R_{\lambda }h=\alpha R_{\lambda }^{(s)}h+\sum_{k=0}^{m_{s}}c_{s.k}(\Psi
_{s,k}-\alpha e^{-i\frac{\sqrt{\lambda -\lambda s,k}}{\varepsilon }t}\varphi
_{s,k}(y/\varepsilon ))-R_{\lambda }h_{1}.  \label{vvv}
\end{equation}
This implies equality (\ref{131}) with $r_{0}=-R_{\lambda }h_{1}.$ Let $%
\Omega _{\varepsilon }^{\prime \prime }$ be obtained from $\Omega
_{\varepsilon }$ by cutting off the parts of channels $C_{j,\varepsilon }$
where $t\geq \frac{7}{8}t_{j}$. Since operator (\ref{b2}) is meromorphic
(due to Theorem \ref{t1}) and has poles of at most first order due to the
Stone formula, $||r_{0}||_{L^{2}(\Omega _{\varepsilon }^{\prime \prime })}$
can be estimated by the right hand side of inequality (\ref{131}). Since $%
(-\varepsilon ^{2}\Delta -\lambda )r_{0}=0$ in $\Omega _{\varepsilon
}^{\prime \prime }$ and $r_{0}=0$ on the lateral side of $\partial \Omega
_{\varepsilon }^{\prime \prime },$ standard a priory estimates for elliptic
equations lead to the estimates on $r_{0}$ in Sobolev norms in $\Omega
_{\varepsilon }^{\prime }.$ These estimates, together with Sobolev imbedding
theorems, justify the estimate (\ref{131}). Similarly, (\ref{132}) follows
from (\ref{vvv}) and Lemma \ref{l22}. This completes the proof of the lemma.

We need two more auxiliary statements in order to prove Theorem \ref{t2}.

\begin{lemma}
\label{tl}Let a real-valued function $f$ belong to $C^{n+1}(R^{1})$ and
\begin{equation}
||f||_{C^{n+1}}=\sum_{k=0}^{n+1}\sup_{x}|f^{(k)}|=A^{+}<\infty ,  \label{s1}
\end{equation}
\begin{equation}
\sum_{k=0}^{n}|f^{(k)}(x)|\geq A^{-}>0,\text{ \ \ \ \ }x\in R^{1}.
\label{s2}
\end{equation}

Then for any $\sigma \leq $ $A^{-}/2,$ the set $\Gamma _{\sigma
}=\{x:|f(x)|\leq \sigma \}$ has the following structure. There exists a
constant $c$ which depends only on $A^{\pm }$ and $n$ and such that, for any
bounded interval $\Delta \subset R^{1},$

a) the number of connected components of $\Gamma _{\sigma }$ in $\Delta $ is
finite and does not exceed $c(|\Delta |+1),$

b) the measure of each connected component of $\Gamma _{\sigma }$ in $\Delta
$ does not exceed $c\sigma ^{1/n}.$
\end{lemma}

\textbf{Remark}. The last estimate can not be improved. In fact, if $%
f(x)=\sin ^{n}x$ then $\Gamma _{\sigma }\cap \lbrack -\frac{\pi }{2},-\frac{%
\pi }{2}]$ $\sim 2\sigma ^{1/n}.$

\textbf{Proof.} We shall denote by $c_{j}$ different constants which depend
on $A^{\pm }$ and $n$ but not on $f.$ If $x\in \Gamma _{\sigma }$ then (\ref
{s2}) implies that
\begin{equation*}
\sum_{k=1}^{n}|f^{(k)}(x)|\geq A^{-}/2,
\end{equation*}
and therefore, $|f^{(k)}(x)|\geq A^{-}/2n$ for the chosen $x$ and some $%
k=k(x),$ $1\leq k\leq n.$ Since $|f^{(k+1)}|\leq $ $A^{+},$ $x\in R^{1},$
there exists an interval $\Delta _{x}$ such that $x\in \Delta _{x},$ $%
|f^{(k)}(x)|\geq A^{-}/4n$ \ on $\Delta _{x},$ and $|\Delta _{x}|=c_{0}=%
\frac{A^{-}}{4nA^{+}}.$ The set of intervals $\Delta _{x}$ covers $\Gamma
_{\sigma }\cap \Delta .$ Hence, one can select a finite number of intervals $%
\Delta _{x}$ covering $\Gamma _{\sigma }\cap \Delta .$ Then one can \ omit
some of them in such a way that the remaining intervals still cover $\Gamma
_{\sigma }\cap \Delta $ with multiplicity at most two. This leaves us with
at most $2(\frac{|\Delta |}{c_{0}}+1)\leq c_{1}(|\Delta |+1)$ intervals $%
\Delta _{x}$ covering $\Gamma _{\sigma }\cap \Delta .$ Thus, it is enough to
prove the lemma for an individual interval $\Delta ^{\prime }$ $\ $(one of
the intervals $\Delta _{x})$ such that $|\Delta ^{\prime }|=c_{0}$ and $%
|f^{(k)}(x)|\geq c_{2}$ \ on $\Delta ^{\prime }$ for some fixed value of $k,$
$1\leq k\leq n.$

Equations $f(x)=\pm \sigma $ have at most $k$ solutions on $\Delta ^{\prime
}.$ In fact, if there exist $k+1$ points where $f(x)=\sigma $ then there are
$k$ \ intermediate points where $f^{\prime }(x)=0.$ Thus, there are $k-1$
points where $f^{\prime \prime }(x)=0,$ and so on. Finally, there has to be
a point where $f^{(k)}(x)=0.$ This contradicts the assumption that $%
|f^{(k)}(x)|\geq c_{2}$ \ on $\Delta ^{\prime }.$ Hence, the set $\Gamma
_{\sigma }\cap \Delta ^{\prime }$ consists of at most $k+1$ intervals. It
remains only to show that the length of these intervals does not exceed $%
c\sigma ^{1/k}.$ In order to estimate this length, we assume that there is
an interval $[x_{1},x_{1}+h]$ where $|f(x)|\leq \sigma ,$ $|f^{(k)}(x)|\geq
c_{2}.$ Put $h^{\prime }=h/k$ and consider the $k$-th difference
\begin{equation}
\Delta _{k}=f(x_{1})-\left(
\begin{array}{c}
k \\
1
\end{array}
\right) f(x_{1}+h^{\prime })+\left(
\begin{array}{c}
k \\
2
\end{array}
\right) f(x_{1}+2h^{\prime })-...+(-1)^{k}f(x_{1}+kh^{\prime }).  \label{ad}
\end{equation}
There exists a point $\xi _{k}\in $ $[x_{1},x_{1}+h]$ such that $\Delta
_{k}=(h^{\prime })^{k}f^{(k)}(\xi _{k}).$ Thus, $|\Delta _{k}|\geq \frac{%
c_{2}h^{k}}{k^{k}}=c_{3}h^{k}.$ On the other hand, from (\ref{ad}) and the
estimate $|f(x)|\leq \sigma $ it follows that $|\Delta _{k}|\leq \sigma
2^{k}.$ Hence, $c_{3}h^{k}\leq \sigma 2^{k},$ i.e. $h\leq c\sigma ^{1/k}.$
This completes the proof of the lemma.

\begin{lemma}
\label{llast}Let a set of functions $f_{\varepsilon }=f_{\varepsilon
}(\lambda ),$ $\varepsilon \rightarrow 0,$\ on a closed interval $l\subset
R^{1},$ have the form
\begin{equation}
f_{\varepsilon }=\sum_{j=1}^{M}C_{j}(\lambda )e^{i\frac{g_{j}(\lambda )}{%
\varepsilon }},  \label{fe}
\end{equation}
where functions $C_{j}(\lambda )$ and real valued, functions $g_{j}(\lambda
) $ are analytic, there are no two functions $g_{j}(\lambda )$ whose
difference is a constant, and
\begin{equation}
\sum_{j=1}^{M}|C_{j}(\lambda )|\geq 1.  \label{nc}
\end{equation}
Then, for any $\eta >0,$ the set
\begin{equation}
\Gamma _{\eta }(\varepsilon )=\{\lambda :|f_{\varepsilon }(\lambda )|\leq
e^{-\eta /\varepsilon }\}  \label{gt}
\end{equation}
is thin (see the definition in the introduction).
\end{lemma}

\textbf{Proof.} Consider the set $\Gamma _{0}$ where $g_{i}^{\prime
}(\lambda )=g_{j}^{\prime }(\lambda )$ for some $i\neq j.$ Due to the
analyticity of the functions $g_{j}(\lambda ),$ this set consists of a
finite number points. Let us denote the number of points in $\Gamma _{0}$ by
$c_{1}.$ Let $\Gamma ^{\delta }$ be the $\delta /2$-neighborhood of $\Gamma
^{0}.$ Then
\begin{equation}
|g_{i}^{\prime }(\lambda )-g_{j}^{\prime }(\lambda )|\geq a(\delta )>0,\text{
\ \ \ }i\neq j,\text{ \ \ \ \ }\lambda \in l\text{ }\backslash \text{ }%
\Gamma ^{\delta }.  \label{vc}
\end{equation}
Consider the functions
\begin{equation*}
\widehat{f}_{\varepsilon }(\mu )=\sum_{j=1}^{M}C_{j}(\varepsilon \mu )e^{i%
\frac{g_{j}(\varepsilon \mu )}{\varepsilon }},\text{ \ \ \ \ }\varepsilon
\mu \in l\text{ }\backslash \text{ }\Gamma ^{\delta }.
\end{equation*}
For any $k,$%
\begin{equation}
\frac{d^{k}}{d\mu ^{k}}\widehat{f}_{\varepsilon }(\mu
)=\sum_{j=1}^{M}[g_{j}^{\prime }(\varepsilon \mu )]^{k}C_{j}(\varepsilon \mu
)e^{i\frac{g_{j}(\varepsilon \mu )}{\varepsilon }}+O(\varepsilon ),
\label{van}
\end{equation}
We move the remainders to the left hand side and consider (\ref{van}) with $%
1\leq k\leq M$ as equations for unknowns $C_{j}(\varepsilon \mu )e^{i\frac{%
g_{j}(\varepsilon \mu )}{\varepsilon }}.$ The matrix of this system of
equations with the elements $a_{j,k}=[g_{j}^{\prime }(\varepsilon \mu )]^{k}$
is a Vandermond matrix, and its determinant is bounded from below due to (%
\ref{vc}). This and (\ref{nc}) imply that
\begin{equation*}
\sum_{j=1}^{M}|\frac{d^{k}}{d\mu ^{k}}\widehat{f}_{\varepsilon }(\mu )|\geq
A^{-}(\delta )>0
\end{equation*}
if $\varepsilon $ is small enough. It also follows from (\ref{van}) that
\begin{equation*}
\sum_{j=1}^{M+1}|\frac{d^{k}}{d\mu ^{k}}\widehat{f}_{\varepsilon }(\mu
)|\leq A^{+}.
\end{equation*}
Hence, Lemma \ref{tl} is applicable to at least one of the functions Re$%
\widehat{f}_{\varepsilon }(\mu )$ or Im$\widehat{f}_{\varepsilon }(\mu )$ on
each connected interval of the set $l$ $\backslash $ $\Gamma ^{\delta }$
stretched by a factor of $\varepsilon ^{-1}$. Since we have at most $c_{1}+1$
those intervals, this implies that the set $\{\lambda :|f_{\varepsilon
}(\lambda )|\leq \sigma \}$ can be covered by $\Gamma ^{\delta }$ and $%
c_{2}(\delta )\varepsilon ^{-1}$ intervals of length $c_{2}(\delta )\sigma
^{1/M}.$ We take $\sigma =e^{-\eta /\varepsilon },$ and this completes the
proof of the lemma.

\textbf{Proof of Theorem \ref{t2}}. The proof is based on a representation
of the resolvent $R_{\lambda }$ of the problem in $\Omega _{\varepsilon }$
through the resolvents $R_{v,\lambda }$ of the operators $H=-\varepsilon
^{2}\Delta $ in the spider domains $\Omega _{v,\varepsilon },$ formed by an
individual junction, which corresponds to a vertex $v$, and all the channels
with an end at this junction, where the channels are extended to infinity if
they have finite length. Let us consider the slices $D_{j,\varepsilon }$ of
the finite channels $C_{j,\varepsilon },$ $j>m$, defined by the conditions $%
t_{j}\leq t\leq t_{j}+3\varepsilon $ where $t_{j}=4l_{j}/5$. We construct
the following partition of the unity on $\Omega _{\varepsilon }:$%
\begin{equation*}
\sum_{v\in V}\phi _{v}=1,
\end{equation*}
where $V$ is the set of all the vertices $v$ of the limiting graph $\Gamma ,$
$\phi _{v}\in C^{\infty }(\Omega _{\varepsilon }),$ and is defined as
follows. The function $\phi _{v}$ is equal to one on the junction $J_{v},$
which corresponds to the vertex $v$, on the infinite channels adjacent to $%
J_{v}$ and on the parts of the finite channels adjacent to $J_{v}$ where $%
t\leq t_{j}+\varepsilon .$ The function $\phi _{v}$ is equal to zero on the
parts of finite channels adjacent to $J_{v}$ where $t\geq t_{j}+2\varepsilon
,$ and also on all the other junctions and channels which are not adjacent
to $J_{v}.$ Let $\psi _{v}\in C^{\infty }(\Omega _{\varepsilon }),$ $\psi
_{v}=1$ on the support of $\phi _{v},$ $\psi _{v}=0$ on the parts of finite
channels adjacent to $J_{v}$ where $t\geq t_{j}+3\varepsilon ,$ and also on
all other junctions and channels which are not adjacent to $J_{v}.$

We fix a vertex $v=\widehat{v}.$ Let $J_{\widehat{v}}$ be the corresponding
junction of $\Omega _{\varepsilon }.$ We choose the parametrization on $%
\Gamma $ in such a way that the value $t=0$ on all the edges adjacent to $%
\widehat{v}$ corresponds to $\widehat{v}.$ The origin ($t=0$) on all the
other edges can be chosen at any end of the edge. We are going to justify
the asymptotic expansion (\ref{psias}) in the domain $C(\widehat{v})$
consisting of the infinite channels adjacent to $J_{\widehat{v}}$ and the
parts $t<3l_{j}/5$ of the finite channels $C_{j,\varepsilon }$ adjacent to $%
J_{\widehat{v}}.$ Moreover, it will be shown that the function $\varsigma $
in the asymptotic expansion satisfies equation (\ref{greq}), conditions (\ref
{uv}) at infinity, and the GC (\ref{gc}). Since $\widehat{v}$ is arbitrary
and the union of all domains $C(\widehat{v}),$ $\widehat{v}\in V$, covers
all the channels, the validity of (\ref{psias}) in $C(\widehat{v})$
justifies the statements of Theorem \ref{t2}.

Let us show that the asymptotic expansion (\ref{psias}) in $C(\widehat{v})$
for any scattering solution $\Psi _{s,k}^{(\varepsilon )}$ follows from a
similar expansion for functions of the form $u=R_{\lambda }f,$ where $f\in
L^{2}(\Omega _{\varepsilon })$ is supported in $\cup D_{j,\varepsilon }$. In
fact, let $u=\psi _{v_{1}}\Psi _{s,k,v_{1}}^{(\varepsilon )},$ where the
vertex $v_{1}=v_{1}(s)$ corresponds to the first junction $J_{v_{1}}$
encountered by the incident wave, $\Psi _{s,k,v_{1}}^{(\varepsilon )}$ is
the solution of the scattering problem in the spider domain $\Omega
_{v_{1},\varepsilon },$ and the function $u$ is considered as a function in $%
\Omega _{\varepsilon }$ which is equal to zero outside of the support of $%
\psi _{v_{1}}.$ Then
\begin{equation*}
(-\varepsilon ^{2}\Delta -\lambda )u=f,\text{ \ \ }f:=-\varepsilon
^{2}[\nabla \psi _{v_{1}}\cdot \nabla \Psi _{s,k,v_{1}}^{(\varepsilon
)}+(\Delta \psi _{v_{1}})\Psi _{s,k,v_{1}}^{(\varepsilon )}].
\end{equation*}
Obviously, $f\in L^{2}(\Omega _{\varepsilon })$ and $f$ is supported in $%
\cup D_{j,\varepsilon }.$ From statement (3) of Theorem \ref{t1} it follows
that there exists the unique outgoing solution $v=R_{\lambda }f$ $\ $of the
equation $(-\varepsilon ^{2}\Delta -\lambda )v=f,$ \ $\lambda \in l$ $%
\backslash $ $\{\lambda ^{j}\}.$ Then
\begin{equation*}
\Psi _{s,k}^{(\varepsilon )}=\psi _{v_{1}}\Psi _{s,k,v_{1}}^{(\varepsilon
)}-R_{\lambda }f,
\end{equation*}
since this function satisfies (\ref{b9}) and (\ref{b10}). From here and
Lemma \ref{l21} it follows that the asymptotic expansion (\ref{psias}) and
the properties of $\varsigma $ mentioned in Theorem \ref{t2} hold for $\Psi
_{s,k}^{(\varepsilon )}$ in $C(\widehat{v})$ if the corresponding properties
are valid for $R_{\lambda }f$ in $C(\widehat{v}).$ Hence, the proof of the
theorem will be complete as soon as we show that, for any $f\in L^{2}(\Omega
_{\varepsilon })$ with the support in $\cup D_{j,\varepsilon },$ the
function $u=R_{\lambda }f$ has expansion (\ref{psias}) in $C(\widehat{v})$
with $\beta _{j,n}=0$ and $\varsigma $ satisfying the GC\ (\ref{gc}).

Consider the operator $P_{\lambda }$ defined by the formula
\begin{equation}
P_{\lambda }h=\sum_{v\in V}\psi _{v}R_{v,\lambda }(\phi _{v}h),\text{ \ \ \ }%
\lambda \in l,  \label{222}
\end{equation}
where $h\in L^{2}(\Omega _{\varepsilon })$ is supported in $\cup
D_{j,\varepsilon },$ $l$ is defined in the statement of Theorem \ref{t2},
and the resolvents $R_{v,\lambda }$ for real $\lambda \in l$ are understood
in the sense of Theorem \ref{t1}. We look for $u=R_{\lambda }f$ \ in the
form of $P_{\lambda }h$ with an unknown $h\in L^{2}(\Omega _{\varepsilon }).$
This leads to the equation (compare with (\ref{a10}), (\ref{a7}))
\begin{equation}
h+F_{\lambda }h=f,\text{ \ \ \ \ }F_{\lambda }h=-\varepsilon ^{2}\sum_{v\in
V}[2\nabla \psi _{v}\cdot \nabla R_{v,\lambda }(\phi _{v}h)+(\Delta \psi
_{v})R_{v,\lambda }(\phi _{v}h)].  \label{110}
\end{equation}
From here, similarly to (\ref{a12}), it follows that
\begin{equation}
R_{\lambda }f=P_{\lambda }(I+F_{\lambda })^{-1}f  \label{111}
\end{equation}
for Im$\lambda >0.$ Similarly to (\ref{a14}), one can show that the operator
\begin{equation}
(I+F_{\lambda })^{-1}:L_{a}^{2}(\Omega _{\varepsilon })\rightarrow
L_{a}^{2}(\Omega _{\varepsilon })  \label{fr}
\end{equation}
admits a meromorphic extension into the lower half plane Im$\lambda \leq 0$
with the branch points at $\lambda =\lambda _{j,n}.$ The only difference is
that now we use operators $R_{v,\lambda }$ instead of $R_{\lambda }^{(j)},$
and $R_{v,\lambda }$ depend meromorphically on $\lambda ,$ while $R_{\lambda
}^{(j)}$ are analytic in $\lambda .$ So, one needs to refer to a version of
an analytic Fredholm theorem where the operator may have poles (with
residues of finite ranks). This version of the theorem can be found in \cite
{BL}, and applications of this theorem to operators similar to (\ref{fr})
can be found in \cite{V1}, \cite{V2}. Hence, formula (\ref{111}) is
established for all complex $\lambda .$

All the operators in (\ref{111}) depend on $\varepsilon .$ The function $%
(I+F_{\lambda })^{-1}f$ is meromorphic in $\lambda $, and its poles depend
on $\varepsilon .$ In order to find a set on the interval $l$ where the
operator $(I+F_{\lambda })^{-1}$ exists and is bounded uniformly in $\lambda
$ we shall use the following reduction of equation (\ref{110}) to a system
of equations where the domains of the operators do not depend on $%
\varepsilon .$

Recall that $f$ is supported in $\cup D_{j,\varepsilon }.$ Formula (\ref{110}%
) for $F_{\lambda }$ implies that the function $F_{\lambda }h$ is also
supported in $\cup D_{j,\varepsilon }.$ Thus, the support of any solution $h$
of (\ref{110}) belongs to $\cup D_{j,\varepsilon }.$ We shall identify
functions $f$ and $h$ with vector functions whose components are the
restrictions of $f$ and $h,$ respectively,\ to individual domains $%
D_{j,\varepsilon },$ $m+1\leq j\leq N.$ Furthermore, we map $%
D_{j,\varepsilon }$ onto $\varepsilon $-independent domain $D_{j}$ by the
transformation $L_{\varepsilon }^{\prime }$ defined by formulas
\begin{equation}
t-t_{j}=\varepsilon \widehat{t},\text{ \ \ \ }y=\varepsilon \widehat{y}.
\label{st}
\end{equation}
This transformation differs from (\ref{te1}) by a shift in $t$ (compare with
(\ref{te})). The vector functions of variables $(\widehat{t},\widehat{y})$
with components from $L^{2}(D_{j})$ defined by $f,$ $h$ will be denoted by $%
f^{\prime }$ and $h^{\prime },$ respectively. Then equation (\ref{110}) can
be written in the form
\begin{equation*}
h^{\prime }+F_{\lambda }^{\prime }h^{\prime }=f^{\prime },
\end{equation*}
where $F_{\lambda }^{\prime }$ is the $[(N-m)\times (N-m)]$-matrix operator
which corresponds to the operator $F_{\lambda }.$ Here $N-m$ is the number
of finite channels in $\Omega _{\varepsilon }.$ Recall that the entries of
the vectors $f^{\prime }$ and $h^{\prime }$ are functions with the domains $%
D_{j}$, which do not depend on $\varepsilon $ (and $\lambda ).$ Our next
goal is to describe how the entries of the matrix $F_{\lambda }^{\prime }$
depend on $\varepsilon $ and $\lambda .$ It will be done using\ (\ref{110}),
where each resolvent $R_{v,\lambda }$ can be specified using (\ref{132}).

The first term in the right hand side of (\ref{132}),
\begin{equation*}
R_{\lambda }^{(s)}:L^{2}(C_{s,\varepsilon }^{\prime })\rightarrow
L^{2}(C_{s,\varepsilon }^{\prime }),
\end{equation*}
depends on $\varepsilon .$ The transformation (\ref{st}) maps $%
C_{s,\varepsilon }^{\prime }$ onto the $\varepsilon $-independent cylinder $%
C_{s}^{\prime }.$ The operator
\begin{equation*}
\widehat{R}_{\lambda }^{(s)}:=L_{\varepsilon }^{\prime }R_{\lambda
}^{(s)}(L_{\varepsilon }^{\prime })^{-1}:L^{2}(C_{s}^{\prime })\rightarrow
L^{2}(C_{s}^{\prime })
\end{equation*}
does not depend on $\varepsilon $ (Lemma \ref{l22})$,$ and it depends
meromorphically on $\lambda .$ Thus, the contributions from the first term
in the right hand side of (\ref{132}) to the entries of the matrix $%
F_{\lambda }^{\prime }$ are operators which are $\varepsilon $-independent
and meromorphic in $\lambda $. The rest of the terms in the right hand side
of (\ref{132}) (other than the remainder) are operators of finite ranks. Due
to Lemma \ref{l23} (see also the formula (\ref{cs}) for $c_{s,k}=c_{s,k}^{-}$%
), the contributions of these terms to the entries of $F_{\lambda }^{\prime
} $ are $\varepsilon $-independent operators which are analytic in $\lambda $
and are of the rank one, multiplied by functions $q_{v;j,n,s,k}$ of the form
\begin{equation}
q_{v;j,n,s,k}(\lambda ,\varepsilon )=e^{i\frac{\alpha _{j}\sqrt{\lambda
-\lambda _{j,n}}+\beta _{s}\sqrt{\lambda -\lambda _{s,k}}}{\varepsilon }}.
\label{qqq}
\end{equation}
Here $\alpha _{j}=t_{j}$ or $\alpha _{j}=l_{j}-t_{j}$ (independently, $\beta
_{s}=t_{s}$ or $\beta _{s}=l_{s}-t_{s}$). Formula (\ref{132}) leads to $%
\alpha _{j}=t_{j},$ $\beta _{s}=t_{s}$ if 1) the channels $C_{j,\varepsilon
} $ and $C_{s,\varepsilon }$ are adjacent to a common junction, which
corresponds to the vertex $v,$ and 2) the parameter $t$ on both channels $%
C_{j,\varepsilon }$ and $C_{s,\varepsilon }$ is introduced in such a way
that $t=0$ at the vertex $v$. Other options in the choice of $\alpha _{j}$
and $\beta _{s}$ correspond to opposite parametrization of the channels $%
C_{j,\varepsilon }$, $C_{s,\varepsilon },$ or both. If $C_{j,\varepsilon }$
and $C_{s,\varepsilon }$ do not have a common junction which corresponds to
the vertex $v$ then $q_{v;j,n,s,k}=0.$ Thus, the matrix operator $F_{\lambda
}^{\prime }$ can be represented in the form
\begin{equation}
F_{\lambda }^{\prime }=F_{\lambda }^{0}+\left[ \sum_{v;n,k}q_{v;j,n,s,k}(%
\lambda ,\varepsilon )F_{\lambda }^{j,n,s,k}\right] _{j,s>m}+R,  \label{fpr}
\end{equation}
where $F_{\lambda }^{0},$ $F_{\lambda }^{j,n,s,k}$ are $\varepsilon $%
-independent operators, $F_{\lambda }^{0}$ is meromorphic in $\lambda ,$
operators $F_{\lambda }^{j,n,s,k}$ are analytic in $\lambda $ and have rank
one, the summation extends over all the verticis $v$ and over $n\in \lbrack
0,m_{j}],$ $k\in \lbrack 0,m_{s}]$. The operator $R=R(\varepsilon ,\lambda )$
corresponds to the remainder term in (\ref{132}) and has the following
estimate
\begin{equation*}
||R||\leq \frac{Ce^{-\gamma /\varepsilon }}{\Pi _{\lambda ^{j}\in l}|\lambda
-\lambda ^{j}|}.
\end{equation*}

Since the analytic Fredholm theorem \cite{BL} is applicable to the operator $%
I+F_{\lambda }^{\prime },$ from (\ref{fpr}) it follows that it is also
applicable to the operator $I+F_{\lambda }^{0}.$ Let $l^{\delta }$ be the $%
\delta /2$-neighborhood of the set consisting of both the poles $\widehat{%
\lambda }^{j}$ of the operator $(I+F_{\lambda }^{0})^{-1}$ located inside $l$
and the points $\lambda ^{j}\in l.$ Then
\begin{equation}
||(I+F_{\lambda }^{0})^{-1}||\leq C(\delta ),\text{ \ \ \ }||R^{\prime
}||\leq C(\delta )e^{-\gamma /\varepsilon },\text{ \ \ \ \ }\lambda \in l%
\text{ }\backslash \text{ }l^{\delta },  \label{ll}
\end{equation}
where
\begin{equation*}
R^{\prime }=R(I+F_{\lambda }^{0})^{-1}.
\end{equation*}
Formula (\ref{fpr}) implies, for $\lambda \in l$ $\backslash $ $l^{\delta },$
\begin{equation}
(I+F_{\lambda }^{\prime })^{-1}=(I+F_{\lambda }^{0})^{-1}\left[
I+qG+R^{\prime }\right] ^{-1},  \label{qwq}
\end{equation}
where $qG$ is the matrix operator with matrix elements $%
\sum_{v;n,k}q_{v;j,n,s,k}(\lambda ,\varepsilon )G_{\lambda }^{j,n,s,k},$ \ $%
N-m<j,s\leq N.$ Here
\begin{equation*}
G_{\lambda }^{j,n,s,k}=F_{\lambda }^{j,n,s,k}(I+F_{\lambda }^{0})^{-1}.
\end{equation*}
The operators $G_{\lambda }^{j,n,s,k}$ are meromorphic in $\lambda $ and
have rank one.

The equation
\begin{equation}
(I+qG)x=g  \label{eq}
\end{equation}
for $x$ can be reduced to an equation in the finite dimensional space $S$
spanned by the ranges of the operators $G_{\lambda }^{j,n,s,k}.$ We fix a
basis in $S$, reduce equation (\ref{eq}) to the algebraic system $A\widehat{x%
}=\widehat{g}$ for coordinates of the projection of $x$ on $S,$ and solve
the system using the Kramer rule. Since functions $q_{v;j,n,s,k}$ are
bounded when $\lambda \in l,$ the procedure described above allows us to
estimate the norm of the operator $(I+qG)^{-1}$ through $|\det^{-1}A|.$

Hence, there exist a polynomial $P=P(q_{v;j,n,s,k})$ of variables $%
q_{v;j,n,s,k}$ which has the following properties. Its coefficients are
meromorphic in $\lambda $ (with poles belonging to the set $\{\widehat{%
\lambda }^{j}\}\cup \{\lambda ^{j}\}$), the polynomial is linear with
respect to each variable $q_{v;j,n,s,k}$ $,$ and is such that
\begin{equation*}
||(I+qG)^{-1}||\leq C|f_{\varepsilon }(\lambda )|^{-1},\text{\ \ \ \ }%
\lambda \in l\text{ }\backslash \text{ }l^{\delta },\text{ \ \ \ }%
f_{\varepsilon }(\lambda ):=1+P(q_{v;j,n,s,k}(\lambda ,\varepsilon )).
\end{equation*}
The function $f_{\varepsilon }(\lambda )$ here has the form (\ref{fe}) with
one of $g_{j}$ identically equal to zero, and the corresponding coefficient $%
C_{j}$ equal to one. The latter implies (\ref{nc}). Thus, Lemma \ref{llast}
can be applied to the function $f_{\varepsilon }(\lambda )$ above on each
connected interval of the set $l$ $\backslash $ $l^{\delta }.$ There are
only finitely many such intervals. Thus, for any $\eta >0$, there exists a
thin set $\Gamma _{\eta }(\varepsilon )$ such that
\begin{equation*}
||(I+qG)^{-1}||\leq Ce^{\eta /\varepsilon },\text{ \ \ \ }\lambda \in l\text{
}\backslash \text{ }\Gamma _{\eta }(\varepsilon ).
\end{equation*}
We choose $\eta <\gamma ,$ where $\gamma $ is defined in (\ref{ll}). Then
\begin{equation*}
||(I+qG+R^{\prime })^{-1}||\leq Ce^{\gamma /2\varepsilon },\text{ \ \ \ }%
\lambda \in l\text{ }\backslash \text{ }\Gamma _{\eta }(\varepsilon ),
\end{equation*}
when $\varepsilon $ is small enough,. A similar estimates holds for operator
(\ref{qwq}):
\begin{equation*}
||(I+F_{\lambda }^{\prime })^{-1}||\leq Ce^{\gamma /2\varepsilon },\text{ \
\ \ }\lambda \in l\text{ }\backslash \text{ }\Gamma _{\eta }(\varepsilon ).
\end{equation*}
Hence, the same estimate is valid for the operator $(I+F_{\lambda })^{-1},$
and from (\ref{222}), (\ref{111}) it follows that
\begin{equation}
R_{\lambda }f=\sum_{v\in V}\psi _{v}R_{v,\lambda }(\phi _{v}h),\text{ \ \ \ }%
\lambda \in l\text{ },  \label{asas}
\end{equation}
where $h\in L^{2}(\Omega _{\varepsilon })$ is supported in $\cup
D_{j,\varepsilon },$ $j>m$, and
\begin{equation}
||h||_{L^{2}(\Omega _{\varepsilon })}\leq Ce^{\gamma /2\varepsilon
}||f||_{L^{2}(\Omega _{\varepsilon })},\ \ \ \lambda \in l\text{ }\backslash
\text{ }\Gamma _{\eta }(\varepsilon ).  \label{asa}
\end{equation}
Relations (\ref{asas}), (\ref{asa}), (\ref{131}), and Lemma \ref{l21}
together provide the asymptotic expansion (\ref{psias}) for $R_{\lambda }f$
needed to complete the proof of Theorem \ref{t2}.

The last result, which we are going to discuss now, concerns the limiting
behavior of the GC as $\lambda $ approaches $\lambda _{0},$ the bottom of
the absolutely continuous spectrum. We assume that the resolvent (\ref{b2})
does not have a pole at $k=\sqrt{\lambda _{0}}.$ Obviously this assumption
holds for generic domains $\Omega _{\varepsilon }.$ Theorem \ref{t1} implies
that this assumption is equivalent to the absence of bounded solutions of
the homogeneous problem (\ref{a1}) with $\lambda =\lambda _{0}.$ Recall that
the scattering matrix (\ref{scm}) depends on $\lambda >\lambda _{0}$ and the
GC (\ref{gc}) depend on both $\lambda >\lambda _{0}$ and $\varepsilon >0.$

\textbf{Proof of Theorem \ref{tlas}}. Consider an infinite channel $C_{\text{%
s}},$ for which $\lambda _{s,0}=\lambda _{0}.$ Let $\Psi
_{s,0}^{(\varepsilon )}$ be the scattering solution which corresponds to the
incident wave
\begin{equation*}
\psi _{inc}=e^{-i\frac{\sqrt{\lambda -\lambda _{s,0}}}{\varepsilon }%
t}\varphi _{s,0}(y/\varepsilon ).
\end{equation*}
Let $\phi _{s}\in C^{\infty }(\Omega _{\varepsilon }),$ $\phi _{s}=1$ in the
channel $C_{\text{s}}$ for $t\geq 2,$ $\phi _{s}=0$ in $C_{s}$ for $t\leq 1$
and outside of $C_{s}.$ We represent $\Psi _{s,0}^{(\varepsilon )}$ in the
form
\begin{equation*}
\Psi _{s,0}^{(\varepsilon )}=\phi _{s}\psi _{inc}+u,\text{ \ \ }\lambda
>\lambda _{0}.
\end{equation*}
Then $u$ is the outgoing solution of the problem
\begin{equation*}
(-\varepsilon ^{2}\Delta -\lambda )u=f,\text{ \ }x\in \Omega _{\varepsilon };%
\text{ \ \ \ }u=0\text{ \ on }\partial \Omega _{\varepsilon },
\end{equation*}
where $f=-\varepsilon ^{2}(\Delta \phi _{s})\psi _{inc}-2\varepsilon
^{2}\nabla \phi _{s}\nabla \psi _{inc}$ has a compact support. Hence, $%
u=R_{\lambda }f.$ From here, the second statement of Theorem \ref{t1}, and
the absence of a pole at $k=\sqrt{\lambda _{0}}$ it follows that $u$, if
considered as an element of $L_{loc}^{2}(\Omega _{\varepsilon }),$ is
analytic in $z=\sqrt{\lambda -\lambda _{0}}$ in a neighborhood of the point $%
z=0.$ Then from standard local a priory estimates for solutions of elliptic
problems it follows that $u$ is analytic, if considered as an element of any
Sobolev space of functions on any bounded part of $\Omega _{\varepsilon }.$
Hence, the restrictions $u_{j}$ of $u$ to cross-sections $t=2$ of infinite
channels $C_{j}$ are analytic in $z.$ Thus, for any infinite channel $C_{j},$
$u$ is an outgoing solution of the problem
\begin{equation}
(-\varepsilon ^{2}\Delta -\lambda )u=0,\text{ \ }x\in \Omega _{\varepsilon
}\cap \{t>2\};\text{ \ \ }u=0\text{ \ on }\partial \Omega _{\varepsilon
}\cap \{t>2\};\text{ \ \ }u|_{t=2}=u_{j}.  \label{llast1}
\end{equation}
Since $u_{j}$ is analytic in $z=\sqrt{\lambda -\lambda _{0}}$ in a
neighborhood of the point $z=0,$ the coefficients $a_{j,n}$ in the
asymptotic expansion (\ref{a2}) for the solution $u$ of (\ref{llast1}) are
analytic in $z.$ This proves the analyticity of the scattering matrix.

From analyticity of $u$ in $z$ and (\ref{llast1}) it also follows that the
scattering solution $\Psi _{s,0}^{(\varepsilon )}$, when $z=0$, is a
solution of the homogeneous problem (\ref{a1}) with $\lambda =\lambda _{0},$
and satisfies (\ref{inf}). Thus $\Psi _{s,0}^{(\varepsilon )}\equiv 0$ when $%
z=0$ due to Theorem \ref{t1}$.$ This implies that $T=-I$ and completes the
proof of the first statement. The second statement of the theorem is an
obvious consequence of the analyticity of $T_{v}$ and (\ref{gc}). This
completes the proof of the theorem.

\textbf{Remarks concerning Theorem \ref{tlas}.} 1) Consider a bounded domain
$\Omega _{\varepsilon }$ with one junction and several channels of finite
length. Let $\Omega _{\varepsilon }^{\prime }$ be a spider type domain which
one gets by extending the channels of $\Omega _{\varepsilon }$ to infinity.
The spectrum of the problem (\ref{a1}) in $\Omega _{\varepsilon }$ is
discrete, and there exists a sequence of eigenvalues which approach $\lambda
_{0}$ as $\varepsilon \rightarrow 0.$ Each of these eigenvalues has the form
\begin{equation}
\lambda _{n}(\varepsilon )=\lambda _{0}+O(\varepsilon ^{2}).  \label{x}
\end{equation}
Theorem \ref{tlas}, concerning the problem in $\Omega _{\varepsilon
}^{\prime }$, can be used to specify the asymptotic behavior (\ref{x}) of
the eigenvalues $\lambda _{n}(\varepsilon ).$ The last statement of the
theorem and (\ref{x}) indicate that, for generic domains $\Omega
_{\varepsilon },$ the asymptotic behavior of $\lambda _{n}(\varepsilon )$ as
$\varepsilon \rightarrow 0$ (when $n$ is fixed or $n\rightarrow \infty $ not
very fast) is the same as for eigenvalues of the corresponding Dirichlet
problem on the limiting graph with the Dirichlet GC at the vertex. This
result will be discussed in more detail elsewhere.

2) Our paper \cite{MV} contains a mistake in the statement of Theorem 5.1
(which is a simplified version of Theorem \ref{tlas} above) about the form
of the GC at the bottom of the absolutely continuous spectrum: $k\rightarrow
0$ has to be replaced there by $k\rightarrow 0,$ $\varepsilon \rightarrow 0.$
The arguments in the last 5 lines of the proof are wrong, but can be easily
corrected with the additional assumption that $\varepsilon \rightarrow 0.$
(Also, the index of summation in the formulas (5.2), (5.4), (5.6) of that
paper must be $n$, not $j$).

\end{document}